\documentclass[a4paper,11pt]{article}
\pdfoutput=1 

\usepackage{jcappub} 

\usepackage[T1]{fontenc} 

\title{\boldmath Color-flavor locked strange stars admixed with mirror dark matter and the observations of compact stars}

\author[a,1]{S.-H. Yang,\note{Corresponding author.}}
\author[b,c]{and C.-M. Pi}

\affiliation[a]{Institute of Astrophysics,\\Central China Normal University,\\Luoyu Road, Hongshan District, Wuhan, China}
\affiliation[b]{School of Physics and Mechanical \& Electrical Engineering,\\Hubei University of Education,\\Hi-Tech 2 Road, East Lake Hi-Tech Zone, Wuhan, China}
\affiliation[c]{Research Center for Astronomy,\\Hubei University of Education,\\Hi-Tech 2 Road, East Lake Hi-Tech Zone, Wuhan, China}

\emailAdd{ysh@ccnu.edu.cn}

\abstract{We investigate the structure and the tidal deformability of the color-flavor locked strange stars admixed with mirror dark matter. Assuming the stars in the GW170817 event have a mirror-dark-matter core or a  mirror-dark-matter halo, the observations of the central compact object within the supernova remnant HESS J1731-347 and the compact objects in the GW190814 and GW170817 events could be explained simultaneously with a pairing gap much smaller than 200 MeV. In contrast, a pairing gap larger than about 200 MeV must be employed without the consideration of a mirror-dark-matter core (halo). More importantly, we find that for the case of the quartic coefficient $a_{4}< 0.589$, if the mass fraction of the mirror mdark matter ($f_{D}$) of the compact stars in GW170817 is in a certain range (eg., $22.8\% < f_{D} < 77.2\%$ for $a_{4}= 0.55$), the minimum allowed value of the pairing gap could be less than 46.5 MeV (i.e., one half of the value of the strange quark mass which is taken as 93 MeV in this paper), which leads to the result that all astrophysical observations mentioned above could be satisfied without violating the conformal bound or the recently proposed positive trace anomally bound.}


\begin{document}
\maketitle
\flushbottom

\section{Introduction}
\label{sec:intro}

Strange stars (SSs) made of strange quark matter (SQM), which consists of up ($u$), down ($d$) and strange ($s$) quarks and electrons, ought to exist in the universe \citep{Farhi1984,Alcock1986,Haensel1986,Alcock1988,Madsen1999,Weber2005}. This is a consequence of the hypothesis proposed in refs.~\cite{Itoh1970,Bodmer1971,Witten1984,Terazawa1989a,Terazawa1989b} that SQM may be the true ground state of baryonic matter. It is also suggested that at sufficiently high density, quarks of different color and flavor form Cooper pairs with the same Fermi momentum, and the ground state of quantum chromodynamics with three flavors is the color-flavor locked (CFL) phase \citep{Alford1999,Alford2001b,Lugones2002,Alford2008}. Thus, SSs might be made of SQM in the CFL phase.

Recently, the mass of the central compact object within the supernova remnant HESS J1731-347 is measured to be $M = 0.77_{-0.17}^{+0.20}\, M_{\odot}$ \citep{Doroshenko2022}. Such a small mass implies that it could be an SS rather than a neutron star (NS) \cite{DiClemente2022,Horvath2023,Oikonomou2023,Chu2023a,Chu2023b,Sagun2023,Rather2023,Das2023,Lopes2024}, since it is less than $1.17\, M_{\odot}$ which is suggested to be the minimum mass of an NS within the supernova remnant \citep{Suwa2018}. What's more, the HESS J1731-347 compact object might be a SS in the CFL phase considering its high temperature \citep{Horvath2023,Sagun2023}.


CFL SSs are also employed to explain the high mass (2.5-2.67 $M_{\odot}$ \citep{Abbott2020}) of GW190814's secondary component \citep{Bombaci2021,Roupas2021,Miao2021,Horvath2021,Zhang2021,Oikonomou2023}, since they could support a larger maximum mass than SSs composed of unpaired SQM \citep{Lugones2003,Horvath2004} \footnote{The compact object in the pulsar binary PSR J0514-4002E with a mass of 2.09 to 2.71 $M_{\odot}$ \citep{Barr2024} could also be a CFL SS in this sense.}. Oikonomou and Moustakidis \citep{Oikonomou2023} found that both the central compact object within the supernova remnant HESS J1731-347 and the compact object in the GW190814 event could be CFL SSs. They also found that these objects can be explained by CFL SSs without violating the conformal bound or the recently proposed positive trace anomally bound ($\left \langle \Theta \right \rangle_{\mu_{B}} \geq 0$ \citep{Fujimoto2022}). However, when they investigate the agreement of the CFL SSs with the tidal deformability observation of GW170817 \citep{Abbott2017,Abbott2018} additionally, it is found that the pairing gap ($\Delta$) must be extremely large ($>$ 200 MeV, such a large pairing gap was also employed in refs. \citep{Roupas2021,Miao2021,Kurkela2024}) and the conformal bound or the positive trace anomally bound must be violated.  

In this paper, we propose an explanation to all of the above-mentioned astrophysical observations which supposes that these compact stars are SSs in the CFL phase, and assumes the stars in GW170817 event are CFL SSs admixed with mirror dark matter (MDM). We will show that in this scenario, the observations of the central compact object within the supernova remnant HESS J1731-347 and the compact objects in GW190814 and GW170817 events could be explained simultaneously without the use of the extremely large pairing gap ($\Delta >$ 200 MeV). More importantly, we find that the conformal bound or the positive trace anomally bound could be satisfied.


Compact stars might contain a dark-matter core or a dark-matter halo made of self-interacting dark matter \cite{Spergel2000,Tulin2018,Bertone2018,Bramante2024}. NSs and SSs admixed with dark matter have been studied extensively \cite{Leung2011,Li2012,Li2012a,Xiang2014,Mukhopadhyay2017,Ellis2018,Ellis2018a,Rezaei2018,Wang2019b,Deliyergiyev2019,Bezares2019,Nelson2019,Garani2019,Das2020,Husain2021,Ivanytskyi2020,Kain2021,
Das2021,Lee2021,Giovanni2021,Berryman2022,Karkevandi2022,Gleason2022,Dengler2022,Lourenco2022a,Lourenco2022b,Leung2022,Das2022a,Das2022b,Miao2022,Rezaei2023,Lenzi2023,Wystub2023,Bauswein2023,Bhattacharya2023,Ruter2023,Routaray2023a,Routaray2023b,Routaray2023c,
Giangrandi2023,Cassing2023,Diedrichs2023,Singh2023,Liu2023,Parmar2023,Cronin2023,Shirke2023,Mariani2024,Hong2024,KanakisPegios2024,Shakeri2024,Vikiaris2024,Thakur2023,Avila2024,Sun2023,Thakur2024,Giangrandi2024,Guha2024,Flores2024,Karkevandi2024,Liu2024,Shirke2024,
Konstantinou2024,Barbat2024,Fibger2024,Shawqi2024,Scordino2024,Thakur2024b,Thakur2024c,Liu2024b,
Mukhopadhyay2016,Panotopoulos2017a,Panotopoulos2017b,Panotopoulos2018b,Sen2022,Jimenez2022,Ferreira2023,Lopes2023,Zhen2024}, and those admixed with MDM have also been studied \cite{Sandin2009,Ciarcelluti2011,Berezhiani2021,Ciancarella2021,Emma2022,Yang2021b,Zollner2022,Zollner2023,Yang2023,Hippert2023}. Technically, this study is an extension of our earlier paper~\citep{Yang2021b} in which SSs (made of unpaired SQM) with a MDM core are investigated. However, this work goes beyond ref.~\citep{Yang2021b} in two ways. First, although the mass and radius of the HESS J1731-347 compact object could be explained without taking into account the pairing effect in the CFL phase, this paper will show that the pairing effect should be considered if one also demands that the maximum mass of the stars is larger than the mass of of GW190814's secondary component (In this paper, we take $M_{\rm max} \geq 2.6\, M_{\odot}$ ). In contrast, $M_{\rm max} \geq 2.08\, M_{\odot}$ is taken in ref.~\citep{Yang2021b} following the observation of PSR J0740-6620 \cite{Cromartie2020,Fonseca2021}, and in that case, all the observations could be explained without the including of the pairing effect. Second, this work investigates not only the SSs with a MDM core, but also the MDM halo scenario, while ref.~\citep{Yang2021b} only focuses on the former. 




In this paper, we will study the structure and the tidal deformability of the CFL SSs admixed with MDM and explain the mass and radius of the HESS J1731-347 compact object, the mass of GW190814's secondary component and the tidal deformability of GW170817 simultaneously. The paper will be arranged as the followings. In section~\ref{EOS}, we briefly review the equation of state (EOS) of the CFL SQM and MDM. In section~\ref{results}, numerical results and discussions are presented. Finally, the summary is given in section~\ref{Summary}.

\section{Equation of state of the CFL SQM and MDM}\label{EOS}
For unpaired SQM, the grand canonical potential per unit volume can be written as \citep{Weissenborn2011,Bhattacharyya2016,Zhou2018,Wang2019}
\begin{equation}
\Omega_{\rm{free}} = \sum_{i=u,d,s,e} \Omega_{i}^{0} + \frac{3 \mu_{b}^4}{4 \pi^2 } (1-a_{4}) +B_{\rm{eff}},
\label{eq:potf}
\end{equation}
where $\Omega_{i}^{0}$ is the grand canonical potential for $u$, $d$, $s$ quarks and electrons described as ideal relativistic Fermi gases. In our calculation, we choose the strange quark mass to be $m_{s}=93$ MeV \citep{Workman2022} while the masses of the up
and down quarks and electrons are set to be zero. The second term at the right hand side of eq.\ (\ref{eq:potf}) accounts for the perturbative quantum chromodynamics (QCD) corrections due to gluon mediated quark interactions to $O(\alpha_{s}^{2})$ \citep{Fraga2001,Alford2005,Weissenborn2011,Bhattacharyya2016}. The quartic coefficient $a_{4}$ represents the degree of quark interaction correction in perturbative QCD, and $a_{4}=1$ corresponds to no QCD corrections.
$\mu_{b}=(\mu_{u}+\mu_{d}+\mu_{s})/3$ is the baryon chemical potential, where $\mu_{i}$ ($i=u,d,s$) are the chemical potentials for each species of quarks. $B_{\rm{eff}}$ is the effective bag constant which includes non-perturbative QCD effects in a phenomenological way.

At large densities, such as in compact star interiors, up, down and strange quarks are assumed to undergo pairing and form the
so-called CFL phase \citep{Alford1999}. For the CFL SQM, the grand canonical potential per unit volume can be written as \citep{Alford2001,Weissenborn2011}
\begin{equation}
\Omega_{\rm{CFL}} = \Omega_{\rm{free}}-\frac{3}{\pi^2} \Delta^2 \mu_{b}^2,
\end{equation}
where $\Delta$ is the pairing gap for the CFL phase. 

In the CFL phase of SQM, the pairing locks the Fermi momenta of all the quarks to a single value, and requires the number densities of
up, down, and strange quarks to be equal \citep{Rajagopal2001}. Thus, when color and electric neutrality are imposed, there are no electrons in the CFL phase. 
The number density of each species of quarks is given by
\begin{equation}
n_{i}=-\frac{\partial\Omega_{\rm{CFL}}}{\partial\mu_{i}},
\end{equation}
which turns out to be
\begin{equation}
n_{u}=n_{d}=n_{s}=\frac{1}{\pi^{2}}[\nu^{3}-(1-a_{4})\mu_{b}^{3}+2\Delta^{2}\mu_{b}],
\end{equation}
where $\nu$ is the common Fermi momentum \cite{Alford2001}
\begin{equation}
\nu=2\mu_{b}-\left(\mu_{b}^{2}+\frac{m_{s}^{2}}{3}\right)^{1/2}.
\end{equation}
The baryon number density is given by
\begin{equation}
n_{b}=\frac{1}{3}(n_{u}+n_{d}+n_{s})=n_{u}=n_{d}=n_{s}.
\end{equation}
The pressure of the CFL SQM is given by
\begin{equation}
p_{Q}=-\Omega_{\rm{CFL}},
\label{eq:pQ}
\end{equation}
and the energy density is obtained from
\begin{equation}
\epsilon_{Q}= \sum_{i=u,d,s}\mu_{i}n_{i}+\Omega_{\rm{CFL}}=3\mu_{b} n_{b}-p_{Q}.
\label{eq:epsQ}
\end{equation}

Mirror dark matter (MDM) is a stable and self-interacting dark matter candidate that emerges from the parity symmetric extension of the Standard Model of particles \cite{Lee1956,Kobzarev1966,Blinnikov1982,Blinnikov1983,Khlopov1991,Foot1991}. For reviews about MDM, see refs.~\cite{Foot2004,Berezhiani2004,BEREZHIANI2005,Okun2007,Foot2014,Berezhiani2018}.

In the minimal parity-symmetric extension of the standard model \cite{Foot1991,Pavsic1974}, the group structure is $G\otimes G$, where $G$ is the gauge group of the standard model. In this model the two sectors are described by the same lagrangians, the only difference between them is that ordinary particles have left-handed interactions, while mirror particles have right-handed interactions. Thus, the microphysics of MDM is the same as that of ordinary matter and we could use the same EOS of the CFL SQM and MDM.

In this study, the CFL SQM is supposed to be the true ground state of baryonic matter. Correspondingly, MDM (more exactly, the CFL mirror SQM consisting of mirror up ($u'$), mirror down ($d'$) and mirror strange ($s'$) quarks) is the ground state of the mirror partner of baryonic matter.
 
The CFL SQM interacts with MDM by gravity. The interactions between quarks and mirror quarks besides gravity have not been studied so far (the neutron–mirror neutron mixing has been studied  \cite{
Berezhiani2006,Berezhiani2009,Goldman2019,McKeen2021,Goldman2022}). However, if such interactions exist, it is reasonable to suppose that they are weak and can be ignored in our study \cite{Berezhiani2021,Yang2021b}.

\section{Results and discussions}\label{results}

For a given EOS of the CFL SQM, the structure of SSs admixed with MDM and their tidal deformabilities are calculated using the two-fluid formalism. The details can be found in our earlier paper~\citep{Yang2021b}. The two-fluid formalism is widely used in the study of the structure and tidal deformability of compact stars with a dark-matter core \cite[e.g.,][]{Ciancarella2021,Yang2021b,Karkevandi2022,Das2022b,Sandin2009,Ciarcelluti2011}. In that formalism, the CFL SQM and MDM do not interact directly, and these two sectors interact only through the gravitational interaction. Following ref.~\citep{Yang2021b}, we define the mass fraction of MDM by $f_{D} \equiv M_{D}/M$, where $M=M_{Q}+ M_{D}$ is the total mass of the star, $M_{Q}$ and $M_{D}$ are the mass of the SQM and the MDM in the star, respectively.

\begin{figure}[tbp]
\centering 
\includegraphics[width=.65\textwidth]{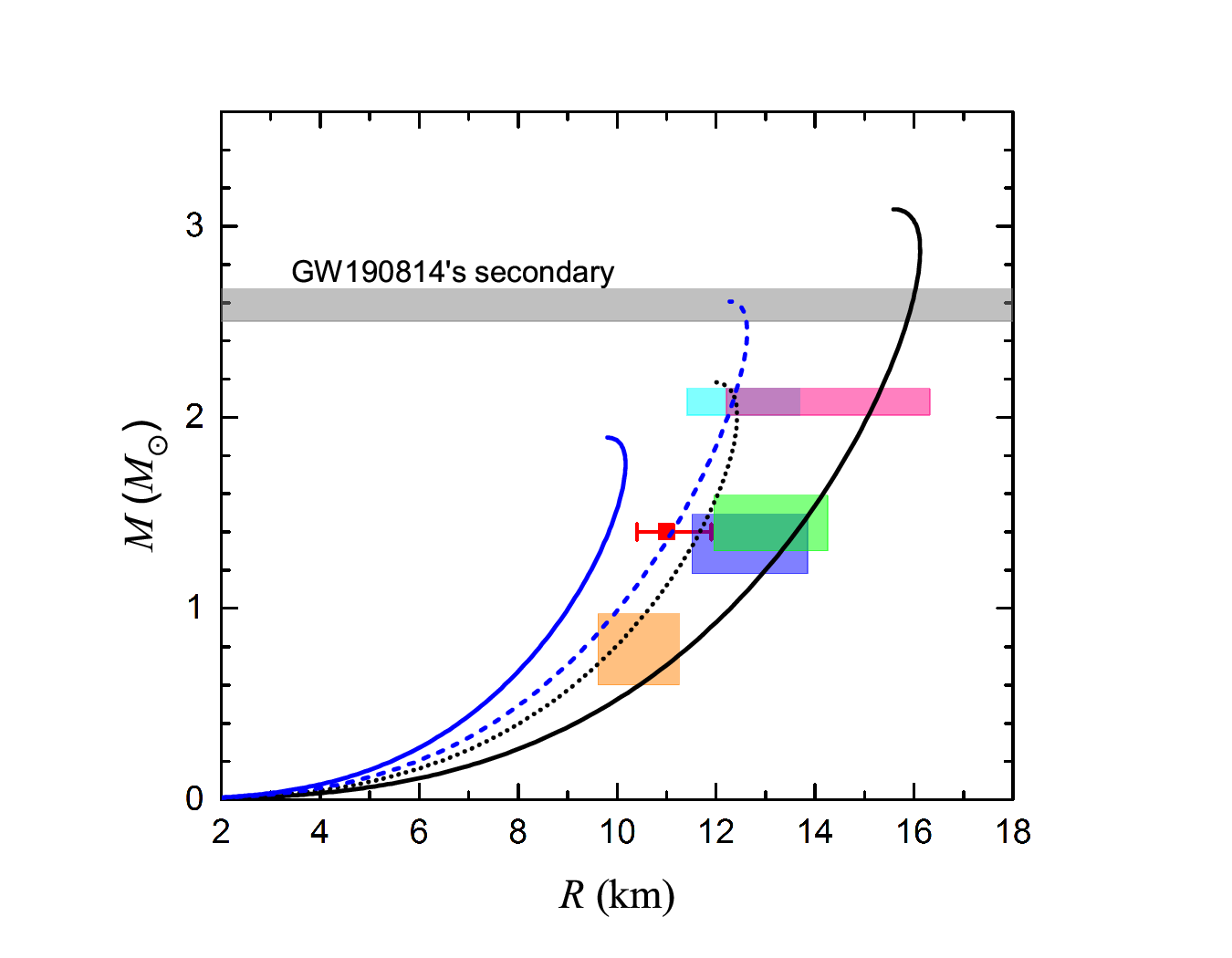}
\caption{\label{rm_delta} The mass-radius relation of pure CFL SSs without MDM (i.e., $f_{D}=0$) for $a_{4}=0.6$. The black curves are for $B_{\rm{eff}}^{1/4}=140$ MeV, and the blue curves are for $B_{\rm{eff}}^{1/4}=170$ MeV. The dotted, solid, dashed lines are for $\Delta=50$, 130, and 220 MeV, respectively. The orange region shows the mass and radius estimates for the central compact object within the supernova remnant HESS J1731-347 \cite{Doroshenko2022}. The grey region shows the mass of GW190814's secondary component \cite{Abbott2020}. The red data is the radius of a $1.4\, M_{\odot}$ compact star constrained by the observations of GW170817 \cite{Capano2020}. The blue and green regions show the mass and radius estimates of PSR J0030+0451 derived from NICER data by Riley et al. \cite{Riley2019} and Miller et al. \cite{Miller2019}, respectively. The cyan and pink regions show the mass of PSR J0740+6620 \cite{Cromartie2020,Fonseca2021}, and the radius of it derived from NICER and XMM-Newton data by Riley et al. \cite{Riley2021} and Miller et al. \cite{Miller2021}, respectively.}
\end{figure}

The mass-radius relation of the CFL SSs without MDM ($f_{D}=0$) is shown in figure~\ref{rm_delta} for $a_{4}=0.6$. All the parameter sets of [$B^{1/4}_{\rm{eff}}$, $\Delta$] are chosen to satisfy both the "2-flavor line" and "3-flavor line" constraints, which will be shown later in figure~\ref{no_MDM}. One can see from figure~\ref{rm_delta} that for a fixed value of $B_{\rm{eff}}^{1/4}$, both the maximum mass and the radius of the $1.4\, M_{\odot}$ star increase as the value of $\Delta$ increases. While, for a fixed value of $\Delta$, both the maximum mass and the radius of the $1.4\, M_{\odot}$ star increase as the value of $B_{\rm{eff}}^{1/4}$ decreases.


\begin{figure}[tbp]
\centering 
\includegraphics[width=.49\textwidth]{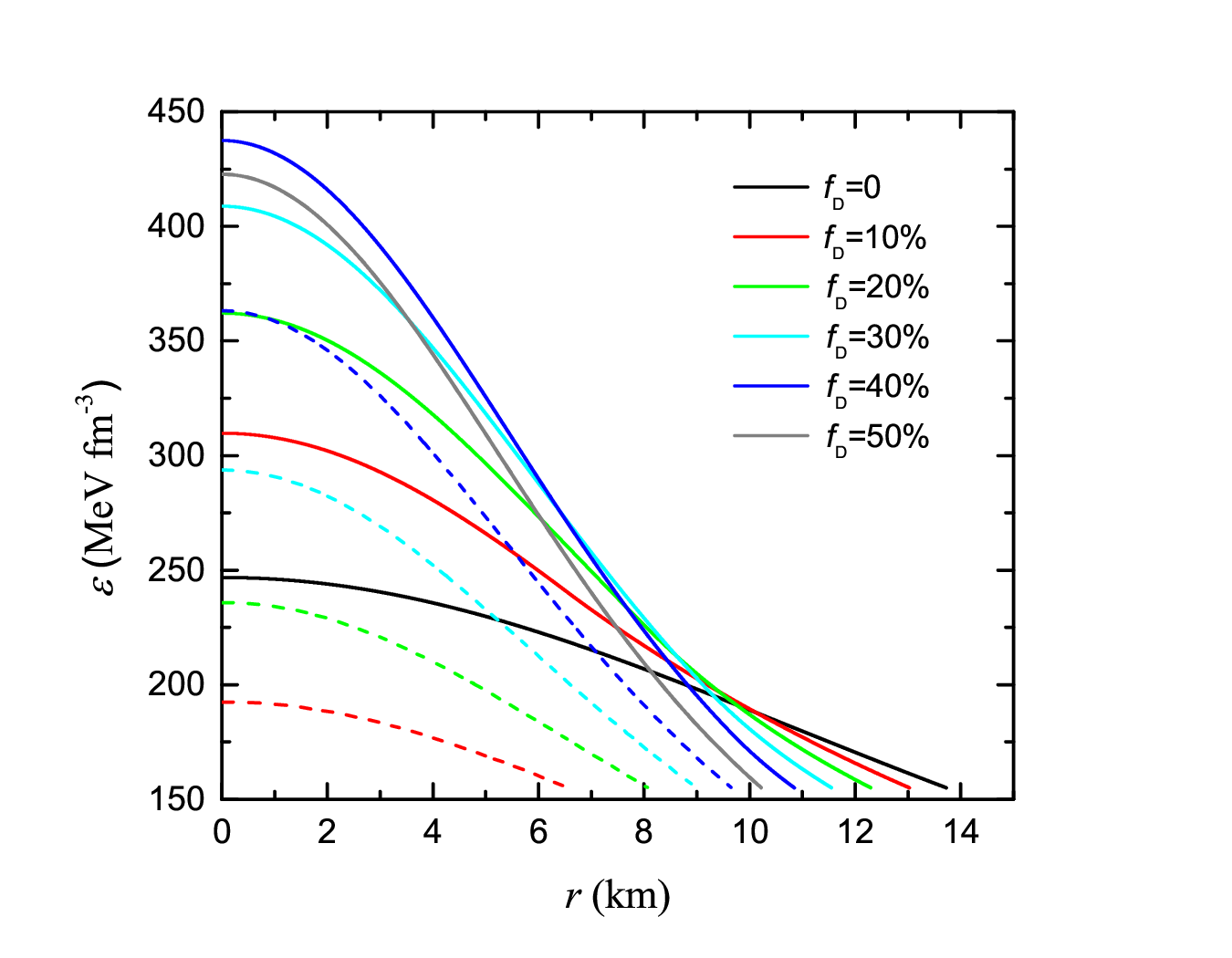}
\hfill
\includegraphics[width=.49\textwidth]{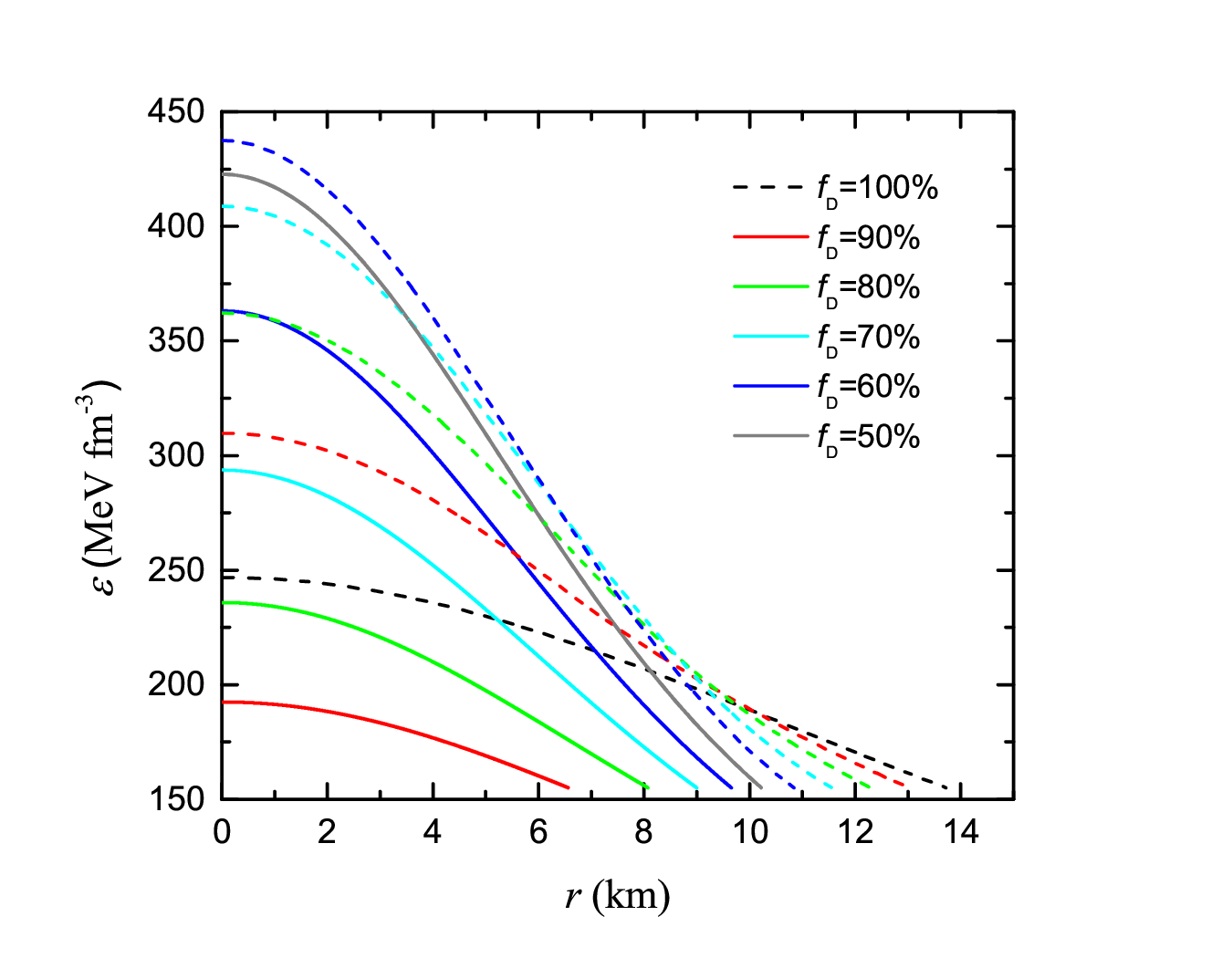}
\caption{\label{rm_ch} Energy density as a function of star radius for MDM admixed CFL SSs with a total mass as $M = 1.8\, M_{\odot}$. The model parameters are chosen to be $a_{4}=0.6$, $B_{\rm{eff}}^{1/4}=140$ MeV, and $\Delta=100$ MeV. The solid and dashed lines are for the SQM and the MDM, respectively (For the case of $f_{D}=$ 50\%, the lines for SQM and MDM totally overlap. Thus, we only show the solid line for $f_{D}=$ 50\%). }
\end{figure}

Figure~\ref{rm_ch} shows the radial profile of the energy density of each component (SQM and MDM) for a $M = 1.8\, M_{\odot}$ star with different values of $f_{D}$. Since the EOS of the SQM and MDM are the same, for the case of $f_{D}=$ 50\% (i.e., $M_{D}= M_{Q}$), the radial profiles of the energy density of SQM and MDM totally overlap, and one has $R_{D}= R_{Q}$ ($R_{Q}$ and $R_{D}$ are the radius of the SQM and the MDM, respectively). One can find the "mirror" behavior of the radial profile of the energy density, e.g., the line for SQM with $f_{D}=20\%$ (the green solid line in the left panel of figure~\ref{rm_ch}) totally overlaps with the line for MDM with $f_{D}=80\%$ (the green dashed line in the right panel of figure~\ref{rm_ch}). For the case of $f_{D}<$ 50\%, one has $R_{D}< R_{Q}$, the MDM forms a dark-matter core (see the left panel of figure~\ref{rm_ch}). While, for the case of $f_{D}>$ 50\%, one has $R_{D}> R_{Q}$, the MDM extends beyond the surface of the SQM and forms a dark-matter halo (see the right panel of figure~\ref{rm_ch}). Mention that $R_{Q}$ is observational radius of the star both for the cases of $R_{D}< R_{Q}$ and $R_{D}> R_{Q}$. One can easily find that $R_{Q}$ decreases with the increasing of $f_{D}$, and its value could be very small. For the case of $f_{D}=100$\%, one has $R_{Q}=0$ and the star is an invisible pure mirror star.

\begin{figure}[tbp]
\centering 
\includegraphics[width=.65\textwidth]{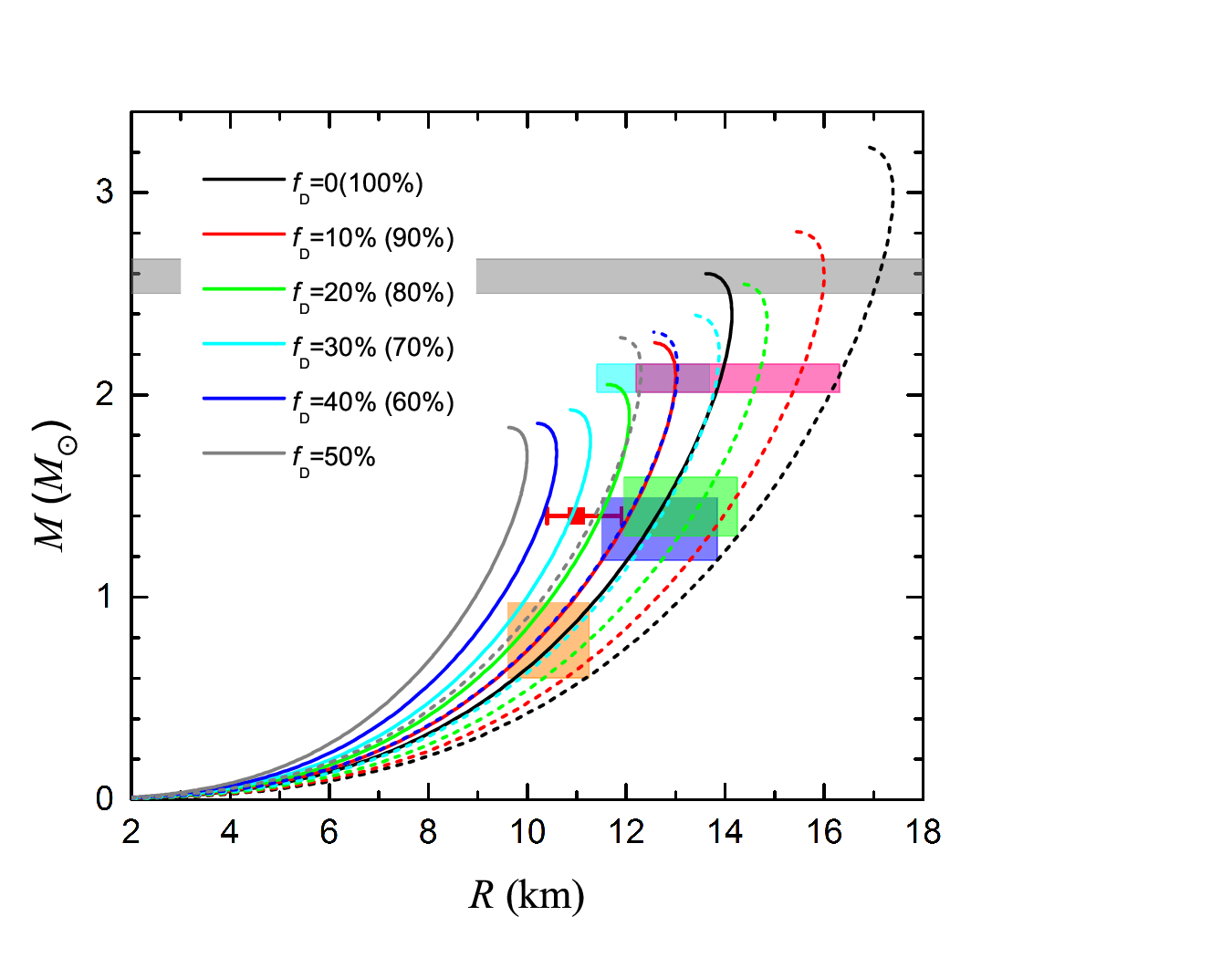}
\caption{\label{rm_f} The mass-radius relation ($R$ is the outermost radius of the star) of MDM admixed CFL SSs for $a_{4}=0.6$ and $\Delta=100$ MeV. The solid and dashed lines are for $B_{\rm{eff}}^{1/4}=141.5$ MeV and $B_{\rm{eff}}^{1/4}=129.4$ MeV, respectively. The data and the shaded regions are the same as these in figure~\ref{rm_delta}.}
\end{figure}

\begin{figure}[tbp]
\centering 
\includegraphics[width=.65\textwidth]{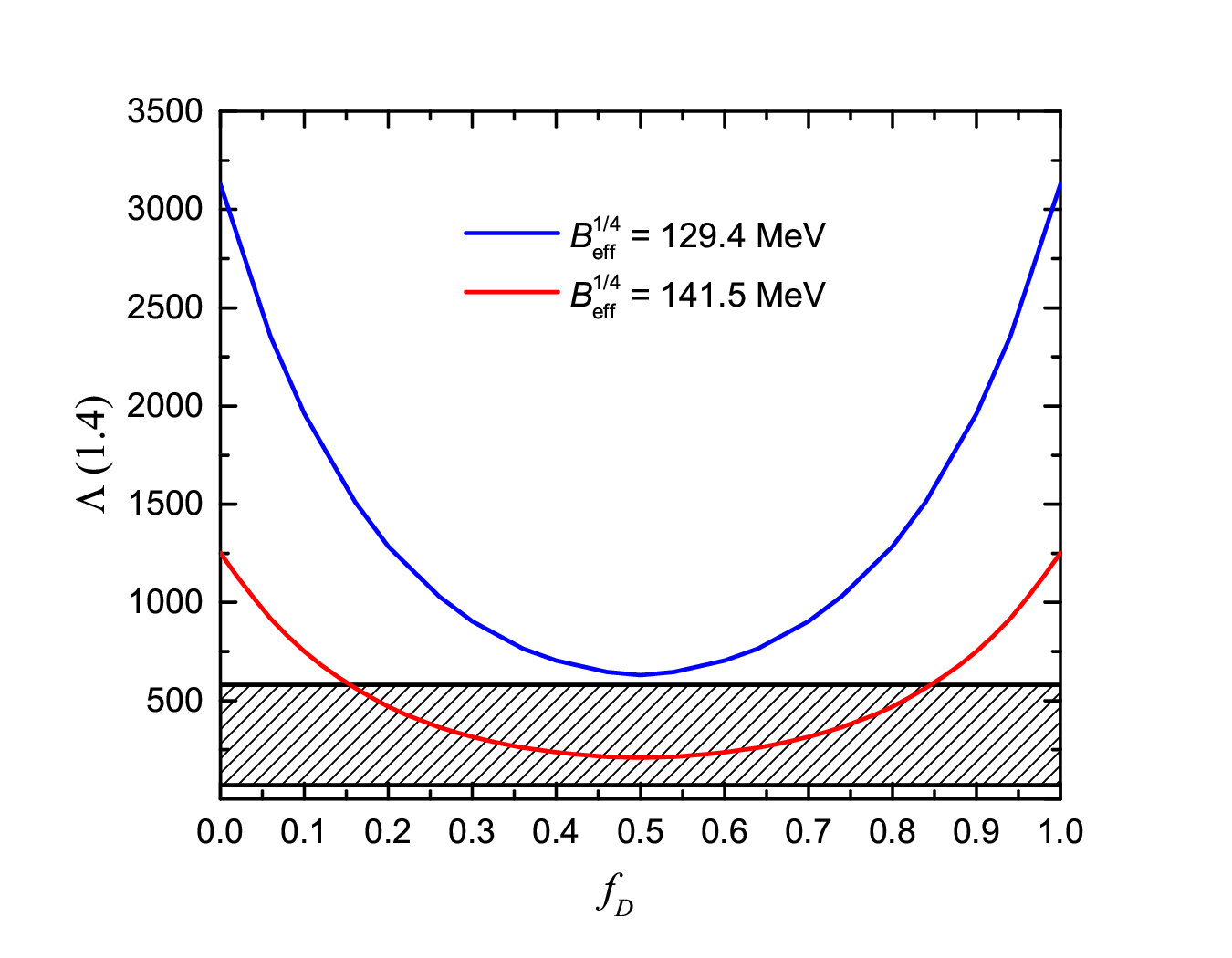}
\caption{\label{lambda_fd} Relation between the dimensionless tidal deformability of a $1.4\, M_{\odot}$ star [$\Lambda(1.4)$] and the mass fraction of MDM ($f_{D}$) for $a_{4}=0.6$ and $\Delta=100$ MeV. The shaded region corresponds to $70<\Lambda(1.4)<580$, which is given by the observation of GW170817.}
\end{figure}

\begin{figure}[tbp]
\centering 
\includegraphics[width=.65\textwidth]{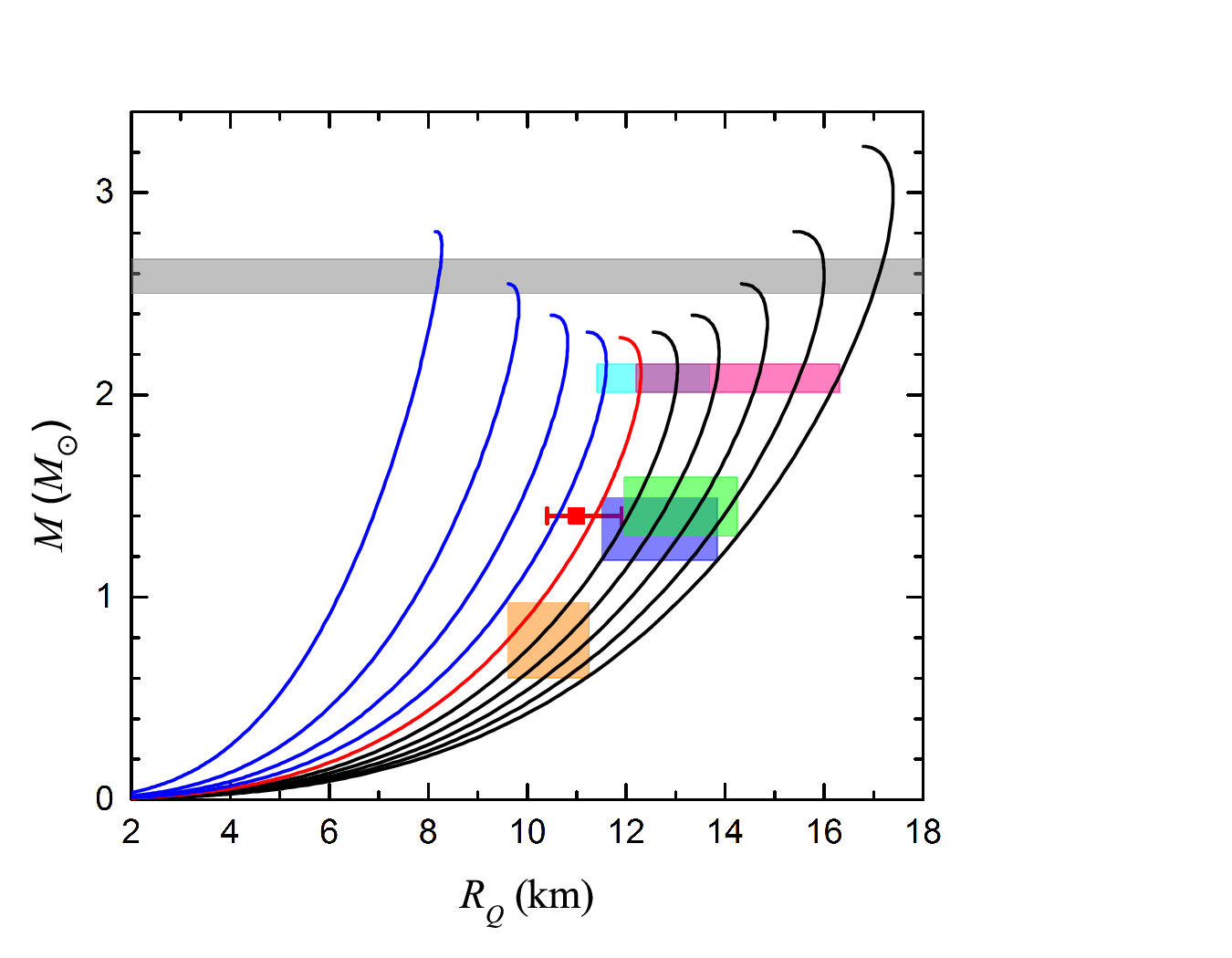}
\caption{\label{rQm_f} The mass-radius relation ($R_{Q}$ is radius of the SQM) of MDM admixed CFL SSs for $a_{4}=0.6$, $\Delta=100$ MeV, and $B_{\rm{eff}}^{1/4}=129.4$ MeV. The red line is for $f_{D}=$ 50\%. From right to left, the black lines are for $f_{D}=$ 0\%, 10\%, 20\%, 30\%, 40\%, and the blue lines are for $f_{D}=$ 60\%, 70\%, 80\%, 90\%. The data and the shaded regions are the same as these in figure~\ref{rm_delta}.}
\end{figure}

Figure~\ref{rm_f} shows the mass-radius relation of the CFL SSs for $a_{4}=0.6$ and $\Delta=100$ MeV with different values of $f_{D}$. In this figure, $R$ is the outermost radius of the star, which equals to $R_{Q}$ for the dark-matter core scenario ($f_{D}<$ 50\%) and equals to $R_{D}$ for the dark-matter halo scenario ($f_{D}>$ 50\%). The value of $B_{\rm{eff}}^{1/4}$ is choosen to be $B_{\rm{eff}}^{1/4}=129.4$ MeV and $B_{\rm{eff}}^{1/4}=141.5$ MeV so that for pure CFL SSs without MDM (i.e., $f_{D}=0$), all the constraints we focus later can be satisfied except the one from the tidal deformability observation of GW170817 (see the black dots in figure~\ref{no_MDM}). One can also see that the maximum mass and the radius of the $1.4\, M_{\odot}$ star decrease with the increasing of $f_{D}$ for the dark-matter core scenario ($f_{D}<$ 50\%). While, for the dark-matter halo scenario ($f_{D}>$ 50\%), both of them increase with the increasing of $f_{D}$. The maximum mass and the radius of the $1.4\, M_{\odot}$ star have a maximum value for $f_{D}=0$ (pure CFL SS) and $f_{D}=100\%$ (pure MDM star), and they have a minimum value for the case of $f_{D}= 50\%$.

Figure~\ref{lambda_fd} shows the relation between the dimensionless tidal deformability of a $1.4\, M_{\odot}$ star [$\Lambda(1.4)$] and the mass fraction of MDM ($f_{D}$) for $a_{4}=0.6$ and $\Delta=100$ MeV. For a fixed value of $B^{1/4}_{\rm{eff}}$, $\Lambda(1.4)$ decreases with the increasing of $f_{D}$ for the case of $f_{D}<$ 50\% (i.e., for the case of the stars with a MDM core). However, it increases with the increasing of $f_{D}$ for $f_{D}>$ 50\% (i.e., for the case of the stars with a MDM halo). Apparently, $\Lambda(1.4)$ reaches the minimum value when $f_{D}=$ 50\%, and it has a maximum value when $f_{D}$ equals to zero (pure CFL SS) or 100\% (pure MDM star). One can also find that $\Lambda(1.4)$ decreases with the increasing of $B^{1/4}_{\rm{eff}}$ for a given value of $f_{D}$. For the case of pure CFL SSs without MDM (i.e., $f_{D}=0$), the tidal deformability of GW170817 cannot be satisfied for both $B_{\rm{eff}}^{1/4}=129.4$ MeV and $B_{\rm{eff}}^{1/4}=141.5$ MeV. For $B_{\rm{eff}}^{1/4}=141.5$ MeV, the tidal deformability of GW170817 can be satisfied if $15.4\%<f_{D}<84.6\%$. However, for $B_{\rm{eff}}^{1/4}=129.4$ MeV, the tidal deformability of GW170817 cannot be satisfied for the entire range of the value of $f_{D}$.

It is worth mentioning that a mirror behavior exists both in figure~\ref{rm_f} and figure~\ref{lambda_fd}. For instance, the curves for $f_{D}=10\%$ and $f_{D}=90\%$ in figure~\ref{rm_f} turn out to be the same (see the red lines). At the same time, one can see from figure~\ref{lambda_fd} that there is no difference between the cases with $f_{D}=10\%$ and $f_{D}=90\%$ since the gravitational radius (i.e., the outermost radius) is important for the tidal deformability parameter. However, these two cases (more exactly, the dark-matter halo scenario and the dark-matter halo scenario) maybe discriminated and the degenercy can be broken when the visible surface of the star is considered (such as the NICER measurements), as shown in figure~\ref{rQm_f}. Different from figure~\ref{rm_f}, the radius in figure~\ref{rQm_f} is the radius of the SQM ($R_{Q}$), which is the visible radius of the star. In figure~\ref{rQm_f}, the lines for the cases with $f_{D}=10\%$ and $f_{D}=90\%$ are quite different, although they remain to have the same maximum mass.

We investigate the allowed parameter space of the CFL SQM model according to the following five constraints \cite[e.g.,][]
{Schaab1997,Weissenborn2011,Pi2015,Zhou2018,Yang2020,Yang2021a,Yang2021b,Pi2022,Oikonomou2023,Ma2023}.

\begin{figure}[tbp]
\centering 
\includegraphics[width=.65\textwidth]{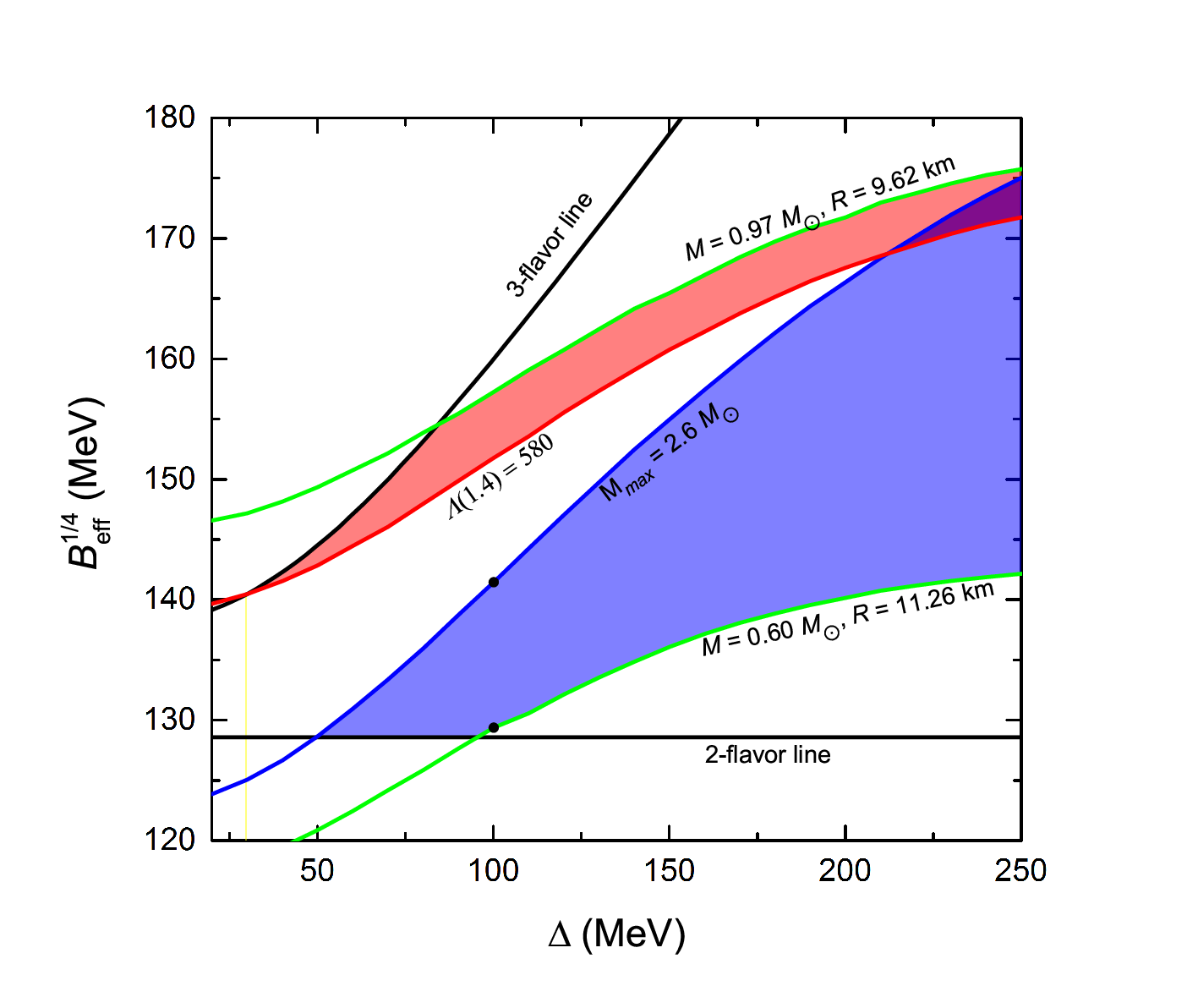}
\caption{\label{no_MDM} Constraints on $B_{\rm{eff}}^{1/4}$ and $\Delta$ for pure CFL SSs without MDM (i.e., $f_{D}=0$) for $a_{4}=0.6$. The blue-shadowed area shows the parameter space which satisfies $M_{\rm max} \geq 2.6\, M_{\odot}$, the red-shadowed area can satisfy $\Lambda(1.4) \leq 580$, and the purple-shadowed area can satisfy both of these constraints. Note that all these shadowed areas can fufill the 2-flavor constaint, the 3-flavor constraint and the HESS J1731-347 constraint. The black dots mark the two parameter sets of [$B^{1/4}_{\rm{eff}}$(MeV), $\Delta$(MeV)] [namely, (129.4, 100) and (141.5, 100)], which are employed for our discussions in figure~\ref{rm_f} and figure~\ref{lambda_fd}.}
\end{figure}

First, the existence of the CFL SSs is based on the idea that the energy per baryon of the CFL SQM is below the energy of the most stable atomic nucleus, $^{56}$Fe ($E/A\sim 930$ MeV) \cite{Witten1984,Zhou2018}. The parameter region that satisfies this constraint is below the 3-flavor line in figure~\ref{no_MDM}.

The second constraint is given by assuming that non-strange quark matter (i.e., two-flavor quark matter made of only $u$ and $d$ quarks)
in bulk has an energy per baryon higher than the one of $^{56}$Fe, plus a 4 MeV correction coming from surface effects
\cite{Farhi1984,Madsen1999,Zhou2018,Yang2021b,Oikonomou2023}. By imposing $E/A\geq 934$ MeV on non-strange quark matter, one ensures that atomic nuclei do not
dissolve into their constituent quarks. The parameter region that satisfies this constraint is above the 2-flavor line in figure~\ref{no_MDM}.

The above two constraints from nuclear structure must be fulfilled when we discuss the constraints from astrophysical observations. In other words, we are only interested in the region between the 3-flavor line and the 2-flavor line in the following.

The third constraint is from the mass and radius measurement of the HESS J1731-347 compact object ($M = 0.77_{-0.17}^{+0.20}\, M_{\odot}$ and $R = 10.4_{-0.78}^{+0.86}$ km \citep{Doroshenko2022}). 
These data are translated into the $B^{1/4}_{\rm{eff}}$--$\Delta$ space in fig.\ \ref{no_MDM}, and the region between the two green lines (the upper one is for [$M\, (M_{\odot}$), $R$\,(km)] set [0.97, 9.62] and the lower one is for [0.6, 11.26]) can fulfill this constraint.

The fourth constraint is that the maximum mass of the CFL SSs
must be greater than the mass of GW190814's secondary component (we take $M_{\rm max} \geq 2.6\, M_{\odot}$ here).
By employing this constraint, the allowed
parameter space is limited to the region below the blue line. The blue-shadowed area in figure~\ref{no_MDM} shows the parameter space which satisfies $M_{\rm max} \geq 2.6\, M_{\odot}$ and the first three constraints mentioned above, namely, the 2-flavor constaint, the 3-flavor constraint and the HESS J1731-347 constraint.

The last constraint is from the tidal deformability observation of GW170817, $\Lambda(1.4)=190 _{-120}^{+390}$, where $\Lambda(1.4)$ is the dimensionless tidal deformability of a $1.4\, M_{\odot}$ star. The parameter region that satisfies this constraint is limited to the area above the red line in figure~\ref{no_MDM}. The red-shadowed area in figure~\ref{no_MDM} shows the parameter space which satisfies $\Lambda(1.4) \leq 580$ and the first three constraints mentioned above (the fourth constraint $M_{\rm max} \geq 2.6\, M_{\odot}$ cannot be satisfied). 

As can be seen from figure~\ref{no_MDM}, the blue-shadowed area overlaps with the red-shadowed area in the purple-shadowed region, which satisfies all the five constraints. We reach to the conclusion that to satisfy all the five constraints, $B^{1/4}_{\rm{eff}}$ must be larger than 168.8 MeV and $\Delta$ must be larger than 212.2 MeV, since the $M_{\rm max} = 2.6\, M_{\odot}$ line cuts across the $\Lambda(1.4) = 580$ line at the point (212.2, 168.8).

\begin{figure}[tbp]
\centering 
\includegraphics[width=.65\textwidth]{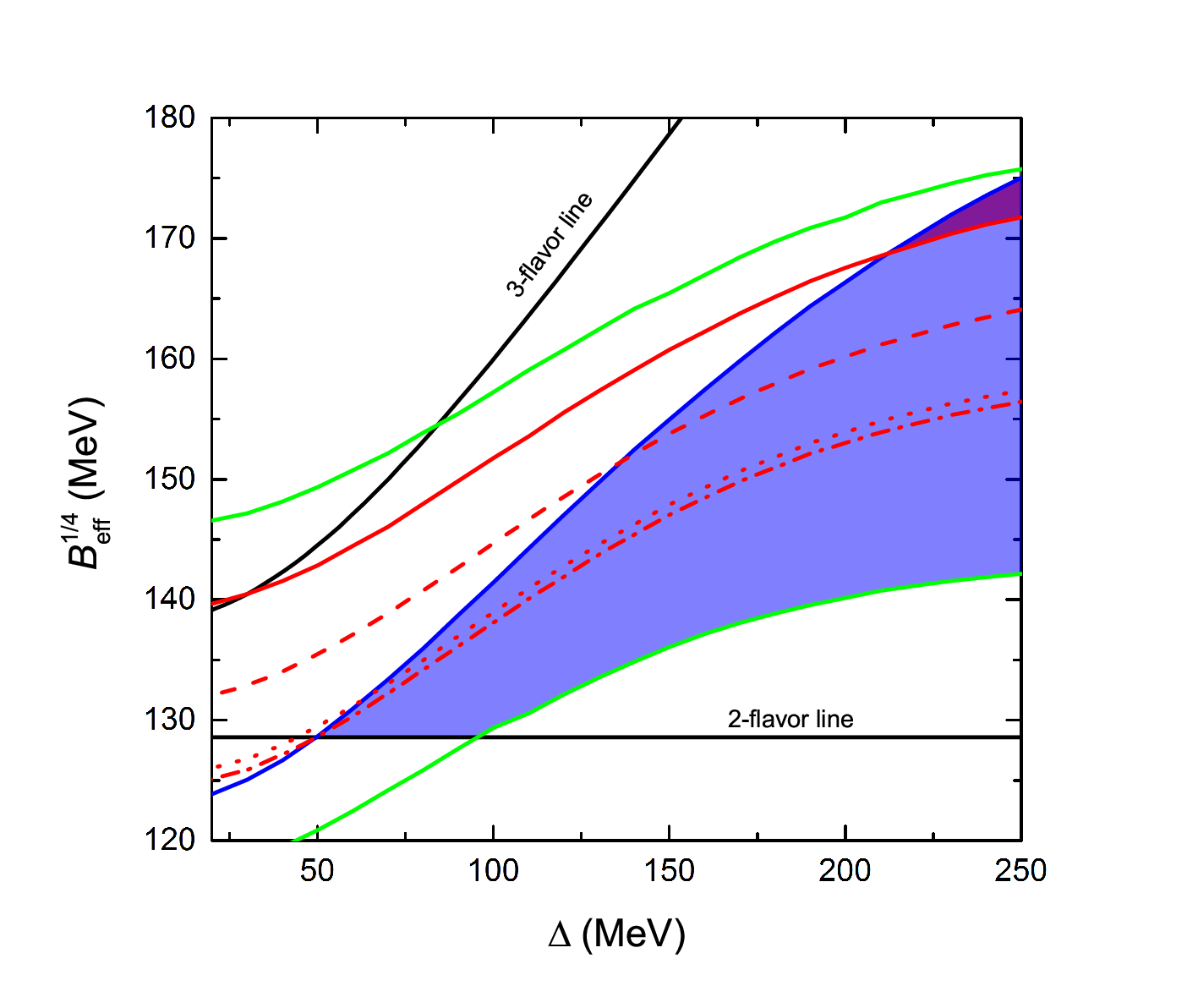}
\caption{\label{cons_a0.6} Constraints on $B_{\rm{eff}}^{1/4}$ and $\Delta$ of the CFL SQM for $a_{4}=0.6$. The only difference from figure~\ref{no_MDM} is that the dashed, dotted and dash-dotted red lines are added, which are for $\Lambda(1.4)=580$ with $f_{D}=$ 10\% (90\%), 20\% (80\%) and 21.6\% (78.4\%), respectively. }
\end{figure}

\begin{figure}[tbp]
\centering 
\includegraphics[width=.65\textwidth]{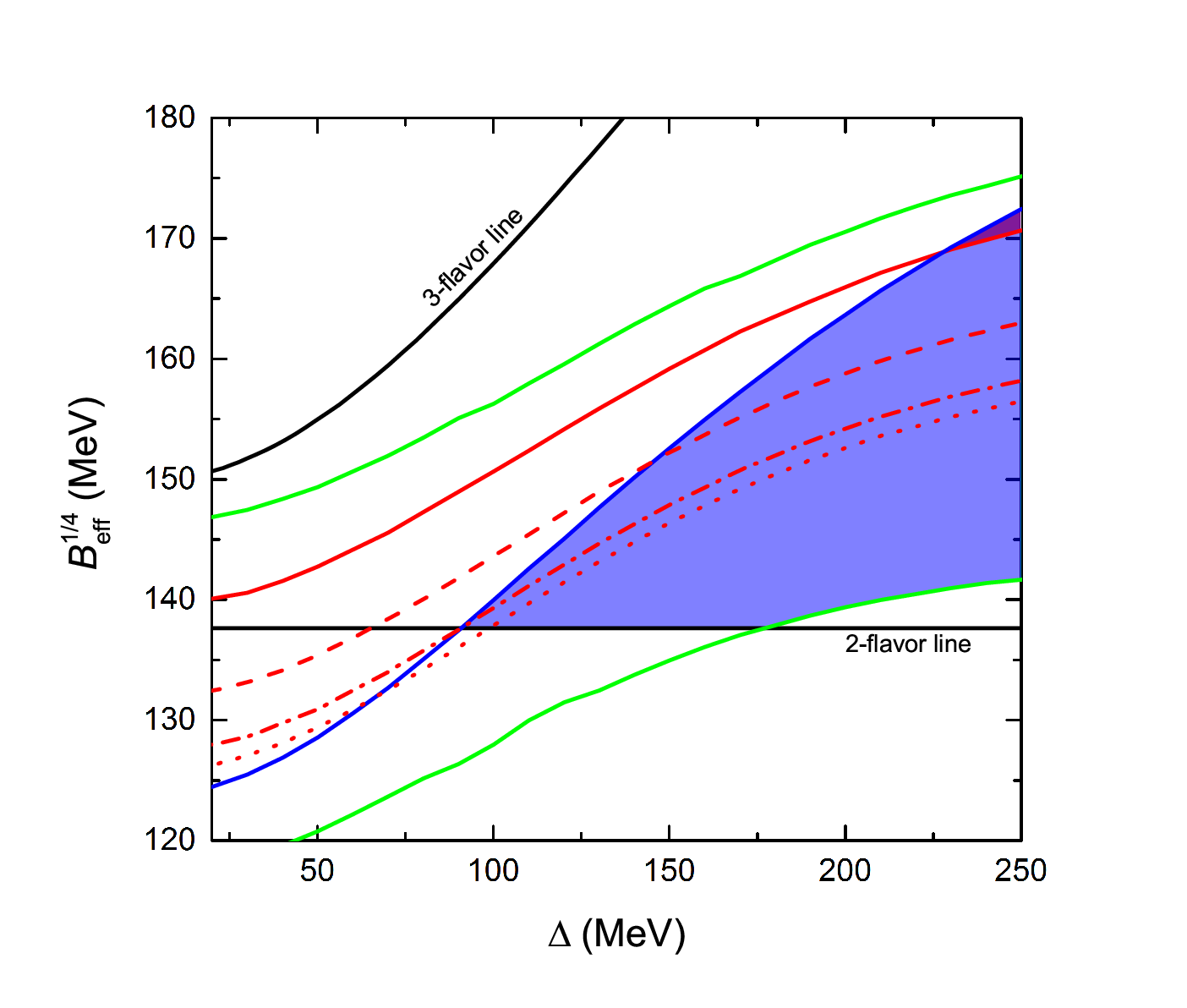}
\caption{\label{cons_a0.8} Similar to figure~\ref{cons_a0.6}, but for $a_{4}=0.8$. The dashed, dotted and dash-dotted red lines are for $\Lambda(1.4)=580$ with $f_{D}=$ 10\% (90\%), 20\% (80\%) and 17.2\% (82.8\%), respectively. }
\end{figure}


Figure~\ref{cons_a0.6} is almost the same as figure~\ref{no_MDM} except that the curves of $\Lambda(1.4)=580$ for the case of  $f_{D}=$ 10\%, 20\% and 21.6\% are presented (These lines overlap with those of  $f_{D}=$ 90\%, 80\% and 78.4\%, respectively).
We stress that the blue-shadowed area in figure~\ref{cons_a0.6} is for the case of pure CFL SSs without MDM ($f_{D}=0$) and satisfies all the five constraints mentioned above except $\Lambda(1.4) \leq 580$. One can see from figure~\ref{cons_a0.6} that as the value of $f_{D}$ increases, the parameter space region which satisfies $\Lambda(1.4) < 580$ (the region above the $\Lambda(1.4)=580$ line) shifts downward. Therefore, assuming GW190814's secondary component and the HESS J1731-347 compact object are pure CFL SSs without MDM, the parameter space which satisfies all the five constraints becomes larger if one assumes the compact stars in GW170817 are MDM admixed CFL SSs with $f_{D}$ in a certain range. It turns out that for the case of $f_{D} < $ 21.6\% and $f_{D} > $ 78.4\%, the minimum values of $B^{1/4}_{\rm{eff}}$ and $\Delta$ needed to satisfy all the five constraints decrease with the increasing of the value of $f_{D}$. For examples, $B^{1/4}_{\rm{eff}} \geq 168.8 $ MeV and $\Delta \geq$ 212.2 MeV are needed for $f_{D}=$ 0; $B^{1/4}_{\rm{eff}} \geq 151.5 $ MeV and $\Delta \geq$ 136.4 MeV are needed for $f_{D}=$ 10\% (90\%); while $B^{1/4}_{\rm{eff}} \geq $ 131.9 MeV and $\Delta \geq$ 63.6 MeV are needed for $f_{D}=$ 20\% (80\%). However, the $f_{D} =$ 21.6\% (78.4\%) line, the $M_{\rm max} = 2.6\, M_{\odot}$ line and the 2-flavor line meet at the point (49.4, 128.6), which means that for the case of $21.6\% \leq f_{D} \leq $ 78.4\%, the minimum allowed values of $B^{1/4}_{\rm{eff}}$ and $\Delta$ are fixed to 128.6 MeV and 49.4 MeV, respectively. We want to stress that in this case, one has the absolute minimum allowed values of $B^{1/4}_{\rm{eff}}$ and $\Delta$. We label them as $B^{1/4}_{\rm{eff,min}}$ and $\Delta_{\rm{min}}$, respectively. Of course, GW190814's secondary component and the HESS J1731-347 compact object could also be MDM admixed CFL SSs. In that case, the blue line and the green lines will shift downword, and the absolute minimum allowed value of $\Delta$ will be a larger one.


Figure~\ref{cons_a0.8} is similar to figure~\ref{cons_a0.6}, but for $a_{4}=0.8$. From this figure, we find that for the scenario that all the stars are pure CFL SSs without MDM, the minimum allowed value of $B^{1/4}_{\rm{eff}}$ and $\Delta$ are 168.9 MeV and 227.6 MeV, respectively. We also find that the $f_{D} =$ 17.2\% (82.8\%) line, the $M_{\rm max} = 2.6\, M_{\odot}$ line and the 2-flavor line meet at the point (90.7, 137.7), which means that if the mass fraction of MDM of the compact stars in GW170817 satisfies $17.2\% \leq f_{D} \leq $ 82.8\% (assuming GW190814's secondary component and the HESS J1731-347 compact object are pure CFL SSs without MDM), the absolute minimum allowed value of $B^{1/4}_{\rm{eff}}$ ($B^{1/4}_{\rm{eff,min}}$) is 137.7 MeV and the absolute minimum allowed value of $\Delta$ ($\Delta_{\rm{min}}$) is 90.7 MeV.

\begin{table}
\centering
\begin{tabular}{|cccc|}
\hline
$a_4$ &  $\Delta_{\rm{min}}$ (MeV) & $B^{1/4}_{\rm{eff,min}}$ (MeV) & $f_{D}$ \\ \hline
0.50  & 14.8  & 123.1 & $24.1\% - 75.9\%$ \\ 
0.55  & 35.4  & 125.9 & $22.8\% - 77.2\%$  \\ 
0.589 & 46.5  & 128.0 & $21.9\% - 78.1\%$ \\
0.60  & 49.4  & 128.6 & $21.6\% - 78.4\%$   \\
0.65  & 61.1  & 131.1 & $20.5\% - 79.5\%$ \\ 
0.70  & 71.5  & 133.4 & $19.4\% - 80.6\%$ \\ 
0.75  & 81.2  & 135.6 & $18.3\% - 81.7\%$ \\ 
0.80  & 90.7  & 137.7 & $17.2\% - 82.8\%$   \\
\hline 
\end{tabular}
\caption{The absolute minimum allowed values of $\Delta$ ($\Delta_{\rm{min}}$) and $B^{1/4}_{\rm{eff}}$ ($B^{1/4}_{\rm{eff,min}}$) for different values of $a_{4}$. The expected range of the value of $f_{D}$ of the compact stars in GW170817 is also presented, which enables $\Delta_{\rm{min}}$ and  $B^{1/4}_{\rm{eff,min}}$ to explain the observations of the central compact object within the supernova remnant HESS J1731-347 and the compact objects in the GW190814 and GW170817 events simultaneously. } 
\label{T1}
\end{table}

We find that 
the absolute minimum allowed values of $B^{1/4}_{\rm{eff}}$ ($B^{1/4}_{\rm{eff,min}}$) and $\Delta$ ($\Delta_{\rm{min}}$) depend on the location of the intersection point of the $M_{\rm max} = 2.6\, M_{\odot}$ line and the 2-flavor line, which changes a lot as different values of $a_{4}$ are employed. The reason is that the location of the 2-flavor line changes significantly although the $M_{\rm max} = 2.6\, M_{\odot}$ line changes a little for different values of $a_{4}$, which can be found by comparing figure~\ref{cons_a0.6} with figure~\ref{cons_a0.8} (This behavior has been clearly shown in figure 1 of ref.~\citep{Bombaci2021}). Details about the change of the absolute minimum allowed values of $B^{1/4}_{\rm{eff}}$ ($B^{1/4}_{\rm{eff,min}}$) and $\Delta$ ($\Delta_{\rm{min}}$) with the value of $a_{4}$ are presented in table~\ref{T1}, where the range of $a_{4}$ is chosen to be $0.5-0.8$ following refs.~\citep{Bhattacharyya2016,Oikonomou2023}. We find that both $B^{1/4}_{\rm{eff,min}}$ and $\Delta_{\rm{min}}$ decrease with the decreasing of the value of $a_{4}$. Table~\ref{T1} also shows the expected range of the value of $f_{D}$ of the compact stars in GW170817 that enables these minimum values of $B^{1/4}_{\rm{eff}}$ and $\Delta$ to explain the observations of the central compact object within the supernova remnant HESS J1731-347 and the compact objects in the GW190814 and GW170817 events simultaneously.


The reduction of $\Delta$ has significant consequences on the conformality of the CFL SQM. Oikonomou and Moustakidis found that whether the conformal bound ($v_{s}\leq c/\sqrt{3}$, where $v_{s}$ is the speed of sound and $c$ is the speed of light) or the recently proposed positive trace anomally bound ($\left \langle \Theta \right \rangle_{\mu_{B}} \geq 0$ \citep{Fujimoto2022}, where $\left \langle \Theta \right \rangle_{\mu_{B}}$ is the trace anomaly at high baryon density) is violated or not depends on the sign of $m^{2}_{s}-4 \Delta^{2}$ \citep{Oikonomou2023}. These bounds will not be violated unless $\Delta < m_{s}/2$ (i.e., $\Delta < 46.5$ MeV, since we take $\Delta = 93$ MeV). From table~\ref{T1}, we find that when $a_{4}< 0.589$, the absolute minimum allowed value of $\Delta$ ($\Delta_{\rm{min}}$) is less than 46.5 MeV (The absolute minimum allowed value of $B^{1/4}_{\rm{eff}}$ ($B^{1/4}_{\rm{eff,min}}$) is less than 128.0 MeV, correspondingly). 
Thus, for $a_{4}< 0.589$, if the compact stars in GW170817 are MDM admixed CFL SSs and $f_{D}$ is in a certain range (eg., $22.8\% < f_{D} < 77.2\%$ for $a_{4}= 0.55$), all of the five constraints could be satisfied without violating the conformal bound or the positive trace anomally bound.

We want to stress that the pairing gap for the CFL phase ($\Delta$) lacks an accurate calculation and it has been investigated in many models \citep{Alford2005,Baym2018,Leonhardt2020}. It is estimated to be in the range of $20 - 250\,\rm{MeV}$ \citep{Kurkela2024}. As shown in figure~\ref{cons_a0.6} and figure~\ref{cons_a0.8}, if MDM is not considered (i.e., for pure CFL SSs), extremely large pairing gap ($\Delta >$ 200 MeV) should be employed to satisfy the observations of compact stars (This agrees with the findings of refs. \citep{Roupas2021,Miao2021,Oikonomou2023}). However, the inclusion of MDM enables us to explain the observations of compact stars with much smaller values of $\Delta$. Especially, the value of $\Delta$ could be less than 46.5 MeV, where the conformal bound or the positive trace anomally bound could also be satisfied. Moreover, there are other motivitions for the reduction of the allowed value of $\Delta$, which were pointed out by Oikonomou and Moustakidis \citep{Oikonomou2023}. First, small $\Delta$ is favored in the explanation of the high surface temperature of the central compact object within the supernova remnant HESS J1731-347 \citep{Horvath2023}. Second, small values of $\Delta$ are accompanied with small $B_{\rm{eff}}$ values (as can be seen from table~\ref{T1}), which favor the existence of CFL SSs more than hybrid stars \citep{Alford2003}.
 

\section{Summary}\label{Summary}

In this paper, we study the structure and the tidal deformability of the CFL SSs admixed with MDM and explain the observations of the central compact object within the supernova remnant HESS J1731-347 and the compact objects in the GW190814 and GW170817 events simultaneously. Our explanation is based on the fact that the mass fraction of MDM ($f_{D}$) of each CFL SS could be different and it depends on the individual history, which was first realized by Ciarcelluti and Sandin \cite{Ciarcelluti2011}. We find that if the CFL SSs in GW170817 are MDM admixed CFL SSs (Assuming GW190814's secondary component and the compact object in HESS J1731-347 are pure CFL SSs without MDM. Our results will not change qualitatively if these stars are also MDM admixed CFL SSs), the observations of these compact objects could be explained simultaneously without the use of the extremely large pairing gap ($\Delta >$ 200 MeV). Moreover, we find that for the case of $a_{4}< 0.589$, if the compact stars in GW170817 are MDM admixed CFL SSs with $f_{D}$ in a certain range, the minimum allowed value of $\Delta$ could be less than 46.5 MeV ($m_{s}/2$), which leads to the result that all of these observations could be satisfied without violating the conformal bound or the positive trace anomally bound.

To avoid the violating the conformal bound or the positive trace anomally bound, the compact stars in GW170817 should be MDM admixed CFL SSs with $f_{D}$ in a certain range (e.g., $22.8\% < f_{D} < 77.2\%$ for $a_{4}= 0.55$). Such high MDM fraction cannot be easily obtained from normal accretion processes \citep{Karkevandi2022}. However, it is reachable through other mechanisms, e.g., a CFL SS could acquire such a large fraction of the MDM if it mergers with a compact star made of MDM \cite{Ciarcelluti2011} (Note that mirror NSs have been studied in ref.~\citep{Hippert2022}. It's reasonable to suppose the existence of mirror SSs made of mirror CFL SQM). Other scenarios of the formation of the high dark matter fractions inside compact stars are presented in refs.~\citep{Karkevandi2022}, which apply equally to our case of MDM inside the CFL SSs.

Finally, we want to stress that although our results are based on the calculation of MDM, they are qualitatively valid for some other kinds of dark matter. The key property used in our approach is that the value of $\Lambda(1.4)$ for MDM admixed CFL SSs could be significantly smaller than that for pure CFL SSs. SSs admixed with some other kinds of dark matter may also have this property. For instance, it is found by Lopes and Das \cite{Lopes2023} that comparing to its value for pure CFL SSs, $\Lambda(1.4)$ could be reduced significantly both for the cases of the bosonic DM (see table 2 in ref. \cite{Lopes2023}) and the fermionic DM (see table 4 in ref. \cite{Lopes2023}). On the other hand, in some circumstances, the value of $\Lambda(1.4)$ for dark matter admixed compact stars is larger than that for pure compact stars, see the upper panel of figure 15 in ref. \cite{Karkevandi2022} for an example. Our approach fails to work in that case.  


\acknowledgments

The authors thank the anonymous referee for his/her helpful comments that greatly improved the quality of the manuscript. This work is supported by the National Key  R\&D Program of China (Grant No.\ 2021YFA0718504) and the Scientific Research Program of the National Natural Science Foundation of China (NSFC, Grant No.\ 12033001).




\begin{thebibliography}{99}



\bibitem{Farhi1984} E. Farhi and R.L. Jaffe, \emph{Strange matter}, \emph{Phys. Rev. D} {\bf 30} (1984) 2379.
\bibitem{Alcock1986} C. Alcock, E. Farhi and A. Olinto, \emph{Strange stars}, \emph{Astrophys. J.} {\bf 310} (1986) 261.
\bibitem{Haensel1986} P. Haensel, J.L. Zdunik and R. Schaefer, \emph{Strange quark stars}, \emph{Astron. Astrophys.} {\bf 160} (1986) 121.
\bibitem{Alcock1988} C. Alcock and A.V. Olinto, \emph{Exotic Phases of Hadronic Matter and their Astrophysical Application}, \emph{Ann. Rev. Nucl. Part. Sci.} {\bf 38} (1988) 161.
\bibitem{Madsen1999} J. Madsen, \emph{Physics and astrophysics of strange quark matter}, \emph{Lect. Notes Phys.} {\bf 516} (1999) 162.
\bibitem{Weber2005} F. Weber, \emph{Strange quark matter and compact stars}, \emph{Prog. Part. Nucl. Phys.} {\bf 54} (2005) 193.
\bibitem{Itoh1970} N. Itoh, \emph{Hydrostatic Equilibrium of Hypothetical Quark Stars}, \emph{Prog. Theor. Phys.} {\bf 44} (1970) 291.
\bibitem{Bodmer1971} A.R. Bodmer, \emph{Collapsed Nuclei}, \emph{Phys. Rev. D} {\bf 4} (1971) 1601.
\bibitem{Witten1984} E. Witten, \emph{Cosmic separation of phases}, \emph{Phys. Rev. D} {\bf 30} (1984) 272.
\bibitem{Terazawa1989a} H. Terazawa, \emph{Super-Hypernuclei in the Quark-Shell Model}, \emph{J. Phys. Soc. Japan} {\bf 58} (1989) 3555.
\bibitem{Terazawa1989b} H. Terazawa, \emph{Super-Hypernuclei in the Quark-Shell Model. II}, \emph{J. Phys. Soc. Japan} {\bf 58} (1989) 4388.
\bibitem{Alford1999} M. Alford, K. Rajagopal and F. Wilczek, \emph{Color-flavor locking and chiral symmetry breaking in high density QCD}, \emph{Nucl. Phys. B} {\bf 537} (1999) 443.
\bibitem{Alford2001b} M. Alford, \emph{Color-Superconducting Quark Matter}, \emph{Ann. Rev. Nucl. Part. Sci.} {\bf 51} (2001) 131. 
\bibitem{Lugones2002} G. Lugones and J. E. Horvath, \emph{Color-flavor locked strange matter}, \emph{Phys. Rev. D} {\bf 66} (2002) 074017. 

\bibitem{Alford2008} M.G. Alford, A. Schmitt, K. Rajagopal and T. Sch\"afer, \emph{Color superconductivity in dense quark matter}, \emph{Rev. Mod. Phys.} {\bf 80} (2008) 1455.
\bibitem{Doroshenko2022} V. Doroshenko, V. Suleimanov, G. P\"uhlhofer and A. Santangelo, \emph{A strangely light neutron star within a supernova remnant}, \emph{Nat. Astron.} {\bf 6} (2022) 1444.
\bibitem{DiClemente2022} F. Di Clemente, A. Drago and G. Pagliara, \emph{Is the compact object associated with HESS J1731-347 a strange quark star? A possible astrophysical scenario for its formation}, \emph{Astrophys. J.} {\bf 967} (2024) 159.
\bibitem{Horvath2023} J.E. Horvath et al., \emph{A light strange star in the remnant HESS J1731-347: Minimal consistency checks}, \emph{Astron. Astrophys.} {\bf 672} (2023) L11.
\bibitem{Sagun2023} V. Sagun, E. Giangrandi, T. Dietrich, O.Ivanytskyi, R. Negreiros and C. Provid\^{e}ncia, \emph{What is the nature of the HESS J1731-347 compact object?}, \emph{Astrophys. J.} {\bf 958} (2023) 49.
\bibitem{Oikonomou2023} P.T. Oikonomou and Ch.C. Moustakidis, \emph{Colour-Flavour Locked Quark Stars in Light of the Compact Object in HESS J1731-347 and the GW190814 Event}, \emph{Phys. Rev. D} {\bf 108} (2023) 063010.
\bibitem{Chu2023a} P.-C. Chu, X.-H. Li, H. Liu, M. Ju and Y. Zhou, \emph{Properties of isospin asymmetric quark matter in quark stars}, \emph{Phys. Rev. C} {\bf 108} (2023) 025808.
\bibitem{Chu2023b} P.-C. Chu, Q. Cao, H. Liu, X.-H. Li, M. Ju, X.-H. Wu and Y. Zhou, \emph{Properties of color-flavor locked matter in a quasiparticle model}, \emph{Eur. Phys. J. C} {\bf 83} (2023) 858.
\bibitem{Rather2023} I.A. Rather, G. Panotopoulos and I. Lopes, \emph{Quark models and radial oscillations: decoding the HESS J1731-347 compact object's equation of state}, \emph{Eur. Phys. J. C} {\bf 83} (2023) 1065.
\bibitem{Das2023} H.C. Das and L.L. Lopes, \emph{Anisotropic strange stars in the spotlight: unveiling constraints through observational data}, \emph{Mon. Not. R. Astron. Soc.} {\bf 525} (2023) 3571.
\bibitem{Lopes2024} L.L. Lopes and H.C. Das, \emph{Spherically symmetric anisotropic strange stars}, \emph{Eur. Phys. J. C} {\bf 84} (2024) 166.

\bibitem{Suwa2018} Y. Suwa, T. Yoshida, M. Shibata, H. Umeda and K. Takahashi,  \emph{On the minimum mass of neutron stars}, \emph{Mon. Not. R. Astron. Soc.} {\bf 482} (2018) 3305. 
\bibitem{Abbott2020} LIGO Scientific and Virgo collaborations, \emph{GW190814: Gravitational Waves from the Coalescence of a 23 Solar Mass Black Hole with a 2.6 Solar Mass Compact Object}, \emph{Astrophys. J. Lett.} {\bf 896} (2020) L44.
\bibitem{Bombaci2021} I. Bombaci, A. Drago, D. Logoteta, G. Pagliara and I. Vida\~na, \emph{Was GW190814 a Black Hole-Strange Quark Star System?}, \emph{Phys. Rev. Lett.} {\bf 126} (2021) 162702.
\bibitem{Roupas2021} Z. Roupas, G. Panotopoulos and I. Lopes, \emph{QCD color superconductivity in compact stars: Color-flavor locked quark star candidate for the gravitational-wave signal GW190814}, \emph{Phys. Rev. D} {\bf 103} (2021) 083015.
\bibitem{Miao2021} Z. Miao, J.-L. Jiang, A. Li and L.-W. Chen, \emph{Bayesian Inference of Strange Star Equation of State Using the GW170817 and GW190425 Data}, \emph{Astrophys. J. Lett.} {\bf 917} (2021) L22.

\bibitem{Horvath2021} J.E. Horvath and P.H.R.S. Moraes, \emph{Modeling a 2.5$M_{\odot}$ compact star with quark matter}, \emph{Int. J. Mod. Phys. D} {\bf 30} (2021) 2150016.
\bibitem{Zhang2021} C. Zhang and R.B. Mann, \emph{Unified interacting quark matter and its astrophysical implications}, \emph{Phys. Rev. D} {\bf 103} (2021) 063018.
\bibitem{Lugones2003} G. Lugones and J.E. Horvath, \emph{High-density QCD pairing in compact star structure}, \emph{Astron. Astrophys.} {\bf 403} (2003) 173.
\bibitem{Horvath2004} J.E. Horvath and G. Lugones, \emph{Self-bound CFL stars in binary systems: Are they ``hidden'' among the black hole candidates?}, \emph{Astron. Astrophys.} {\bf 422} (2004) L1.
\bibitem{Barr2024} E.D. Barr et al., \emph{A pulsar in a binary with a compact object in the mass gap between neutron stars and black holes}, \emph{Science} {\bf 383} (2024) 275.
\bibitem{Fujimoto2022} Y. Fujimoto, K. Fukushima, L.D. McLerran and M. Prasza\l owicz, \emph{Trace Anomaly as Signature of Conformality in Neutron Stars}, \emph{Phys. Rev. Lett.} {\bf 129} (2022) 252702.
\bibitem{Abbott2017} LIGO Scientific and Virgo collaborations, \emph{GW170817: Observation of Gravitational Waves from a Binary Neutron Star Inspiral}, \emph{Phys. Rev. Lett.} {\bf 119} (2017) 161101.
\bibitem{Abbott2018} LIGO Scientific and Virgo collaborations, \emph{GW170817: Measurements of Neutron Star Radii and Equation of State}, \emph{Phys. Rev. Lett.} {\bf 121} (2018) 161101.
\bibitem{Kurkela2024} A. Kurkela, K. Rajagopal and R. Steinhorst, \emph{Astrophysical Equation-of-State Constraints on the Color-Superconducting Gap}, \emph{Phys. Rev. Lett.} {\bf 132} (2024) 262701.

\bibitem{Spergel2000} D.N. Spergel and P.J. Steinhardt, \emph{Observational Evidence for Self-Interacting Cold Dark Matter}, \emph{Phys. Rev. Lett.} {\bf 84} (2000) 3760.
\bibitem{Tulin2018} S. Tulin and H.-B. Yu, \emph{Dark matter self-interactions and small scale structure}, \emph{Phys. Rep.} {\bf 730} (2018) 1.
\bibitem{Bertone2018} G. Bertone and T.M.P. Tait, \emph{A new era in the search for dark matter}, \emph{Nature} {\bf 562} (2018) 51.
\bibitem{Bramante2024} J. Bramante and N. Raj, \emph{Dark matter in compact stars}, \emph{Phys. Rep.} {\bf 1052} (2024) 1.

\bibitem{Leung2011} S.-C. Leung, M.-C. Chu and L.-M. Lin, \emph{Dark-matter admixed neutron stars}, \emph{Phys. Rev. D} {\bf 84} (2011) 107301.
\bibitem{Li2012a} A. Li, F. Huang and R.-X. Xu, \emph{Too massive neutron stars: The role of dark matter?}, \emph{Astropart. Phys.} {\bf 37} (2012) 70.
\bibitem{Li2012} X.Y. Li, F.Y. Wang and K.S. Cheng, \emph{Gravitational effects of condensate dark matter on compact stellar objects}, \emph{JCAP} {\bf 10} (2012) 031.
\bibitem{Xiang2014} Q.-F. Xiang, W.-Z. Jiang, D.-R. Zhang and R.-Y. Yang, \emph{Effects of fermionic dark matter on properties of neutron stars}, \emph{Phys. Rev. C} {\bf 89} (2014) 025803.
\bibitem{Mukhopadhyay2017} S. Mukhopadhyay, D. Atta, K. Imam, D.N. Basu and C. Samanta, \emph{Compact bifluid hybrid stars: hadronic matter mixed with self-interacting fermionic asymmetric dark matter}, \emph{Eur. Phys. J. C} {\bf 77} (2017) 440.
\bibitem{Ellis2018} J. Ellis, G. Hutsi, K. Kannike, L. Marzola, M. Raidal and V. Vaskonen, \emph{Dark matter effects on neutron star properties}, \emph{Phys. Rev. D} {\bf 97} (2018) 123007.
\bibitem{Ellis2018a} J. Ellis, A. Hektor, G. Hutsi, K. Kannike, L. Marzola, M. Raidal and V. Vaskonen, \emph{Search for dark matter effects on gravitational signals from neutron star mergers}, \emph{Phys. Lett. B} {\bf 781} (2018) 607.
\bibitem{Rezaei2018} Z. Rezaei, \emph{Neutron stars with spin polarized self-interacting dark matter}, \emph{Astropart. Phys.} {\bf 101} (2018) 1.
\bibitem{Wang2019b} X.D. Wang, B. Qi, G.L. Yang, N.B. Zhang and S.Y. Wang, \emph{Possible maximum mass of dark matter existing in compact stars based on the self-interacting fermionic model}, \emph{Int. J. Mod. Phys. D} {\bf 28} (2019) 1950148.

\bibitem{Nelson2019} A. Nelson, S. Reddy and D. Zhou, \emph{Dark halos around neutron stars and gravitational waves}, \emph{JCAP} {\bf 07} (2019) 012.
\bibitem{Garani2019} R. Garani, Y. Genolini and T. Hambye, \emph{New analysis of neutron star constraints on asymmetric dark matter}, \emph{JCAP} {\bf 05} (2019) 035.
\bibitem{Deliyergiyev2019} M. Deliyergiyev, A. DelPopolo, L. Tolos, M. LeDelliou, X. Lee and F. Burgio, \emph{Dark compact objects: An extensive overview}, \emph{Phys. Rev. D} {\bf 99} (2019) 063015.
\bibitem{Bezares2019} M. Bezares, D. Vigano and C. Palenzuela, \emph{Gravitational wave signatures of dark matter cores in binary neutron star mergers by using numerical simulations}, \emph{Phys. Rev. D} {\bf 100} (2019) 044049.

\bibitem{Ivanytskyi2020} O. Ivanytskyi, V. Sagun and I. Lopes, \emph{Neutron stars: New constraints on asymmetric dark matter}, \emph{Phys. Rev. D} {\bf 102} (2020) 063028.

\bibitem{Das2020} H.C. Das et al., \emph{Effects of dark matter on the nuclear and neutron star matter}, \emph{Mon. Not. R. Astron. Soc.} {\bf 495} (2020) 4893. 

\bibitem{Kain2021} B. Kain, \emph{Dark matter admixed neutron stars}, \emph{Phys. Rev. D} {\bf 103} (2021) 043009.
\bibitem{Husain2021} W. Husain and A.W. Thomas, \emph{Possible nature of dark matter}, \emph{JCAP} {\bf 10} (2021) 086.
\bibitem{Das2021} H.C. Das, A. Kumar and S.K. Patra, \emph{Dark matter admixed neutron star as a possible compact component in the GW190814 merger event}, \emph{Phys. Rev. D} {\bf 104} (2021) 063028.
\bibitem{Lee2021} B.K.K. Lee, M.-C. Chu and L.-M. Lin, \emph{Could the GW190814 Secondary Component Be a Bosonic Dark Matter Admixed Compact Star?}, \emph{Astrophys. J.} {\bf 922} (2021) 242.

\bibitem{Berryman2022} J.M. Berryman, S. Gardner and M. Zakeri, \emph{Neutron Stars with Baryon Number Violation, Probing Dark Sectors}, \emph{Symmetry} {\bf 14} (2022) 518. 

\bibitem{Karkevandi2022} D. Rafiei Karkevandi, S. Shakeri, V. Sagun and O. Ivanytskyi, \emph{Bosonic dark matter in neutron stars and its effect on gravitational wave signal}, \emph{Phys. Rev. D} {\bf 105} (2022) 023001.

\bibitem{Lourenco2022a} O. Louren\c{c}o, T. Frederico and M. Dutra, \emph{Dark matter component in hadronic models with short-range correlations}, \emph{Phys. Rev. D} {\bf 105} (2022) 023008. 
\bibitem{Gleason2022} T. Gleason, B. Brown and B. Kain, \emph{Dynamical evolution of dark matter admixed neutron stars}, \emph{Phys. Rev. D} {\bf 105} (2022) 023010.

\bibitem{Dengler2022} Y. Dengler, J. Schaffner-Bielich and L. Tolos, \emph{Second Love number of dark compact planets and neutron stars with dark matter}, \emph{Phys. Rev. D} {\bf 105} (2022) 043013.
\bibitem{Giovanni2021} F. Di Giovanni, N. Sanchis-Gual, P. Cerd\'a-Dur\'an and J.A. Font, \emph{Can fermion-boson stars reconcile multimessenger observations of compact stars?}, \emph{Phys. Rev. D} {\bf 105} (2022) 063005.
\bibitem{Leung2022} K.-L. Leung, M.-C. Chu and L.-M. Lin, \emph{Tidal deformability of dark matter admixed neutron stars}, \emph{Phys. Rev. D} {\bf 105} (2022) 123010.

\bibitem{Das2022b} A. Das, T. Malik and A.C. Nayak, \emph{Dark matter admixed neutron star properties in light of gravitational wave observations: A two fluid approach}, \emph{Phys. Rev. D} {\bf 105} (2022) 123034.
\bibitem{Lourenco2022b}O. Louren\c{c}o, C.H. Lenzi, T. Frederico and M. Dutra, \emph{Dark matter effects on tidal deformabilities and moment of inertia in a hadronic model with short-range correlations}, \emph{Phys. Rev. D} {\bf 106} (2022) 043010. 

\bibitem{Das2022a} H.C. Das, A. Kumar, B. Kumar and S.K. Patra, \emph{Dark Matter Effects on the Compact Star Properties}, \emph{Galaxies} {\bf 10} (2022) 14.

\bibitem{Miao2022} Z. Miao, Y. Zhu, A. Li and F. Huang, \emph{Dark Matter Admixed Neutron Star Properties in the Light of X-Ray Pulse Profile Observations}, \emph{Astrophys. J.} {\bf 936} (2022) 69.

\bibitem{Rezaei2023} Z. Rezaei, \emph{Fuzzy dark matter in relativistic stars}, \emph{Mon. Not. R. Astron. Soc.} {\bf 524} (2023) 2015. 
\bibitem{Wystub2023} S. Wystub, Y. Dengler, J.-E. Christian and J. Schaffner-Bielich, \emph{Constraining exotic compact stars composed of bosonic and fermionic dark matter with gravitational wave events}, \emph{Mon. Not. R. Astron. Soc.} {\bf 521} (2023) 1393.

\bibitem{Lenzi2023} C.H. Lenzi, M. Dutra, O. Louren\c{c}o, L.L. Lopes and D.P. Menezes, \emph{Dark matter effects on hybrid star properties}, \emph{Eur. Phys. J. C} {\bf 83} (2023) 266.

\bibitem{Routaray2023a} P. Routaray, A. Quddus, K. Chakravarti and B. Kumar, \emph{Probing the impact of WIMP dark matter on universal relations, GW170817 posterior, and radial oscillations}, \emph{Mon. Not. R. Astron. Soc.} {\bf 525} (2023) 5492. 

\bibitem{Cassing2023} M. Cassing, A. Brisebois, M. Azeem and J. Schaffner-Bielich, \emph{Exotic Compact Objects with Two Dark Matter Fluids}, \emph{Astrophys. J.} {\bf 944} (2023) 130.
\bibitem{Giangrandi2023} E. Giangrandi, V. Sagun, O. Ivanytskyi, C. Provid\^{e}ncia and T. Dietrich, \emph{The Effects of Self-interacting Bosonic Dark Matter on Neutron Star Properties}, \emph{Astrophys. J.} {\bf 953} (2023) 115.

\bibitem{Bauswein2023} A. Bauswein, G. Guo, J.-H. Lien, Y.-H. Lin and M.-R. Wu, \emph{Compact dark objects in neutron star mergers}, \emph{Phys. Rev. D} {\bf 107} (2023) 083002.
\bibitem{Singh2023} D. Singh, A. Gupta, E. Berti, S. Reddy and B.S. Sathyaprakash, \emph{Constraining properties of asymmetric dark matter candidates from gravitational-wave observations}, \emph{Phys. Rev. D} {\bf 107} (2023) 083037.
\bibitem{Diedrichs2023} R.F. Diedrichs, N. Becker, C. Jockel, J.-E. Christian, L. Sagunski and J. Schaffner-Bielich, \emph{Tidal deformability of fermion-boson stars: Neutron stars admixed with ultralight dark matter}, \emph{Phys. Rev. D} {\bf 108} (2023) 064009.    
\bibitem{Parmar2023} V. Parmar, H.C. Das, M.K. Sharma and S.K. Patra, \emph{Influence of dark matter on magnetized neutron stars}, \emph{Phys. Rev. D} {\bf 108} (2023) 083003. 
\bibitem{Cronin2023} J. Cronin, X. Zhang and B. Kain, \emph{Rotating dark matter admixed neutron stars}, \emph{Phys. Rev. D} {\bf 108} (2023) 103016. 
\bibitem{Ruter2023} H.R. R\"{u}ter, V. Sagun, W. Tichy and T. Dietrich, \emph{Quasiequilibrium configurations of binary systems of dark matter admixed neutron stars}, \emph{Phys. Rev. D} {\bf 108} (2023) 124080.  
\bibitem{Bhattacharya2023}S. Bhattacharya, B. Dasgupta, R. Laha and A. Ray, \emph{Can LIGO Detect Nonannihilating Dark Matter?}, \emph{Phys. Rev. Lett.} {\bf 131} (2023) 091401.  

\bibitem{Routaray2023c} P. Routaray et al., \emph{Investigating dark matter-admixed neutron stars with NITR equation of state in light of PSR J0952-0607}, \emph{JCAP} {\bf 10} (2023) 073.
\bibitem{Liu2023} H.-M. Liu, J.-B. Wei, Z.-H. Li, G.F. Burgio and H.-J. Schulze, \emph{Dark matter effects on the properties of neutron stars: optical radii}, \emph{Phys. Dark Universe} {\bf 42} (2023) 101338. 

\bibitem{Shirke2023} S. Shirke, S. Ghosh, D. Chatterjee, L. Sagunski and J. Schaffner-Bielich, \emph{R-modes as a new probe of dark matter in neutron stars}, \emph{JCAP} {\bf 12} (2023) 008.

\bibitem{Routaray2023b} P. Routaray, H.C. Das, J.A. Pattnaik and B. Kumar, \emph{Dark Matter Admixed Neutron Star in the light of HESS J1731-347 and PSR J0952-0607}, arXiv:2307.12748. 

\bibitem{Thakur2024} P. Thakur, T. Malik and T.K. Jha, \emph{Towards Uncovering Dark Matter Effects on Neutron Star Properties: A Machine Learning Approach}, \emph{Particles} {\bf 7} (2024) 80.

\bibitem{Giangrandi2024} E. Giangrandi, A. \'{A}vila, V. Sagun, O. Ivanytskyi and C. Provid\^{e}ncia, \emph{The impact of asymmetric dark matter on the thermal evolution of nucleonic and hyperonic compact stars}, \emph{Particles} {\bf 7} (2024) 179. 

\bibitem{Karkevandi2024} D. Rafiei Karkevandi, M. Shahrbaf, S. Shakeri and S. Typel, \emph{Exploring the distribution and impact of bosonic dark matter in neutron stars}, \emph{Particles} {\bf 7} (2024) 201.  

\bibitem{Mariani2024} M. Mariani, C. Albertus, M. del Rosario Alessandroni, M.G. Orsaria, M.A. P\'{e}rez-Garc\'{i}a and I.F. Ranea-Sandoval, \emph{ Constraining self-interacting fermionic dark matter in admixed neutron stars using multimessenger astronomy}, \emph{Mon. Not. R. Astron. Soc.} {\bf 527} (2024) 6795.  
\bibitem{Avila2024} A. \'{A}vila, E. Giangrandi, V. Sagun, O. Ivanytskyi and C. Provid\^{e}ncia, \emph{Rapid neutron star cooling triggered by dark matter}, \emph{Mon. Not. R. Astron. Soc.} {\bf 528} (2024) 6319.

\bibitem{Hong2024} B. Hong and Z. Ren, \emph{Mixed dark matter models for the peculiar compact object in remnant HESS J1731-347 and their implications for gravitational wave properties}, \emph{Phys. Rev. D} {\bf 109} (2024) 023002.

\bibitem{KanakisPegios2024} A. Kanakis-Pegios, V. Petousis, M. Veselský, J. Leja and Ch.C. Moustakidis, \emph{Constraints for the X17 boson from compact objects observations}, \emph{Phys. Rev. D} {\bf 109} (2024) 043028. 

\bibitem{Shakeri2024} S. Shakeri and D. Rafiei Karkevandi, \emph{Bosonic dark matter in light of the NICER precise mass-radius measurements}, \emph{Phys. Rev. D} {\bf 109} (2024) 043029. 

\bibitem{Thakur2023}P. Thakur, T. Malik, A. Das, T.K. Jha and C. Provid\^{e}ncia, \emph{Exploring robust correlations between fermionic dark matter model parameters and neutron star properties: A two-fluid perspective}, \emph{Phys. Rev. D} {\bf 109} (2024) 043030.

\bibitem{Guha2024} A. Guha and D. Sen, \emph{Constraining the mass of fermionic dark matter from its feeble interaction with hadronic matter via dark mediators in neutron stars}, \emph{Phys. Rev. D} {\bf 109} (2024) 043038. 

\bibitem{Flores2024} C.V. Flores, C.H. Lenzi, M. Dutra, O. Louren\c{c}o and J.D.V. Arba\~{n}il, \emph{Gravitational wave asteroseismology of dark matter hadronic stars}, \emph{Phys. Rev. D} {\bf 109} (2024) 083021.

\bibitem{Vikiaris2024} M. Vikiaris, V. Petousis, M. Veselsk\'{y} and Ch. C. Moustakidis, \emph{Supramassive dark objects with neutron star origin}, \emph{Phys. Rev. D} {\bf 109} (2024) 123006.

\bibitem{Sun2023} H. Sun and D. Wen, \emph{A new criterion for the existence of dark matter in neutron stars}, \emph{Phys. Rev. D} {\bf 109} (2024) 123037.

\bibitem{Konstantinou2024} A. Konstantinou, \emph{The Effect of a Dark Matter Core on the Structure of a Rotating Neutron Star}, \emph{Astrophys. J.} {\bf 968} (2024) 83.
 
\bibitem{Barbat2024} M.F. Barbat, J. Schaffner-Bielich and L. Tolos, \emph{A comprehensive study of compact stars with dark matter}, \emph{Phys. Rev. D} {\bf 110} (2024) 023013.
\bibitem{Liu2024} H.-M. Liu, J.-B. Wei, Z.-H. Li, G.F. Burgio, H.C. Das and H.-J. Schulze, \emph{Dark matter effects on the properties of neutron stars: compactness and tidal deformability}, \emph{Phys. Rev. D} {\bf 110} (2024) 023024.

\bibitem{Fibger2024} M. Fibger, R. Negreiros, O. Louren\c{c}o and M. Dutra, \emph{Cooling of hadronic stars with dark matter components}, \emph{J. Phys. G: Nucl. Part. Phys.} {\bf 51} (2024) 095202. 

\bibitem{Shirke2024} S. Shirke, B.K. Pradhan, D. Chatterjee, L. Sagunski and J. Schaffner-Bielich, \emph{Effects of Dark Matter on f-mode oscillations of Neutron Stars}, arXiv:2403.18740.

\bibitem{Scordino2024} D. Scordino and I. Bombaci, \emph{Dark matter admixed neutron stars with a realistic nuclear equation of state from chiral nuclear interactions}, arXiv:2405.19251.

\bibitem{Shawqi2024} S. Shawqi and S.M. Morsink, \emph{Interpreting Mass and Radius Measurements of Neutron Stars with Dark Matter Halos}, arXiv:2406.03332. 

\bibitem{Thakur2024b} P. Thakur, A. Kumar, V.B. Thapa, V. Parmar and M. Sinha, \emph{Exploring non-radial oscillation modes in dark matter admixed neutron stars}, arXiv:2406.07470.  

 \bibitem{Thakur2024c} P. Thakur, T. Malik, A. Das, T.K. Jha, B.K. Sharma and C. Provid\^{e}ncia, \emph{Feasibility of dark matter admixed neutron star based on recent observational constraints}, arXiv:2408.03780.
\bibitem{Liu2024b} Y. Liu, H.-B. Li, Y. Gao, L. Shao and Z. Hu, \emph{Effects from Dark Matter Halos on X-ray Pulsar Pulse Profiles}, arXiv:2408.04425.

\bibitem{Mukhopadhyay2016} P. Mukhopadhyay and J. Schaffner-Bielich, \emph{Quark stars admixed with dark matter}, \emph{Phys. Rev. D} {\bf 93} (2016) 083009.
\bibitem{Panotopoulos2017a} G. Panotopoulos and I. Lopes, \emph{Gravitational effects of condensed dark matter on strange stars}, \emph{Phys. Rev. D} {\bf 96} (2017) 023002.

\bibitem{Panotopoulos2017b} G. Panotopoulos and I. Lopes, \emph{Radial oscillations of strange quark stars admixed with condensed dark matter}, \emph{Phys. Rev. D} {\bf 96} (2017) 083013. 


\bibitem{Panotopoulos2018b} G. Panotopoulos and I. Lopes, \emph{Radial oscillations of strange quark stars admixed with fermionic dark matter}, \emph{Phys. Rev. D} {\bf 98} (2018) 083001. 
 
\bibitem{Sen2022} D. Sen and A. Guha, \emph{Vector dark boson mediated feeble interaction between fermionic dark matter and strange quark matter in quark stars}, \emph{Mon. Not. R. Astron. Soc.} {\bf 517} (2022) 518.

\bibitem{Jimenez2022} J.C. Jim\'{e}nez and E.S. Fraga, \emph{Radial Oscillations of Quark Stars Admixed with Dark Matter}, \emph{Universe} {\bf 8} (2022) 34.
\bibitem{Ferreira2023} O. Ferreira and E.S. Fraga, \emph{Strange magnetars admixed with fermionic dark matter}, \emph{JCAP} {\bf 04} (2023) 012.
\bibitem{Lopes2023} L.L. Lopes and H.C. Das, \emph{Strange stars within bosonic and fermionic admixed dark matter}, \emph{JCAP} {\bf 05} (2023) 034.
\bibitem{Zhen2024} Y. Zhen, T.-T. Sun, J.-B. Wei, Z.-Y. Zheng and H. Chen, \emph{Radial Oscillations of Strange Quark Stars Admixed with Dark Matter}, \emph{Symmetry} {\bf 16} (2024) 807. 

\bibitem{Sandin2009} F. Sandin and P. Ciarcelluti, \emph{Effects of mirror dark matter on neutron stars}, \emph{Astropart. Phys.} {\bf 32} (2009) 278.
\bibitem{Ciarcelluti2011} P. Ciarcelluti and F. Sandin, \emph{Have neutron stars a dark matter core?}, \emph{Phys. Lett. B} {\bf 695} (2011) 19.
\bibitem{Ciancarella2021} R. Ciancarella, F. Pannarale, A. Addazi and A. Marciano, \emph{Constraining mirror dark matter inside neutron stars}, \emph{Phys. Dark Universe} {\bf 32} (2021) 100796.
\bibitem{Berezhiani2021} Z. Berezhiani, R. Biondi, M. Mannarelli and F. Tonelli, \emph{Neutron-mirror neutron mixing and neutron stars}, \emph{Eur. Phys. J. C} {\bf 81} (2021) 1036.
\bibitem{Yang2021b} S.-H. Yang, C.-M. Pi and X.-P. Zheng, \emph{Strange stars with a mirror-dark-matter core confronting with the observations of compact stars}, \emph{Phys. Rev. D} {\bf 104} (2021) 083016.
\bibitem{Emma2022} M. Emma, F. Schianchi, F. Pannarale, V. Sagun and T. Dietrich, \emph{Numerical Simulations of Dark Matter Admixed Neutron Star Binaries}, \emph{Particles} {\bf 5} (2022) 273. 
\bibitem{Zollner2022} R. Z\"ollner and B. K\"ampfer, \emph{Exotic Cores with and without Dark-Matter Admixtures in Compact Stars}, \emph{Astronomy} {\bf 1} (2022) 36.
\bibitem{Zollner2023} R. Z\"ollner, M. Ding and B. K\"ampfer, \emph{Masses of Compact (Neutron) Stars with Distinguished Cores}, \emph{Particles} {\bf 6} (2023) 217. 

\bibitem{Yang2023} S. Yang, C. Pi, X. Zheng and F. Weber, \emph{Confronting Strange Stars with Compact-Star Observations and New Physics}, \emph{Universe} {\bf 9} (2023) 202.
\bibitem{Hippert2023} M. Hippert, E. Dillingham, H. Tan, D. Curtin, J. Noronha-Hostler and N. Yunes, \emph{Dark Matter or Regular Matter in Neutron Stars? How to tell the difference from the coalescence of compact objects}, \emph{Phys. Rev. D} {\bf 107} (2023) 115028. 


\bibitem{Cromartie2020} H.T. Cromartie et al., \emph{Relativistic Shapiro delay measurements of an extremely massive millisecond pulsar}, \emph{Nat. Astron.} {\bf 4} (2020) 72.
\bibitem{Fonseca2021} E. Fonseca et al., \emph{Refined Mass and Geometric Measurements of the High-Mass PSR J0740+6620}, \emph{Astrophys. J. Lett.} {\bf 915} (2021) L12.

\bibitem{Weissenborn2011} S. Weissenborn, I. Sagert, G. Pagliara, M. Hempe and J. Schaffner-Bielich, \emph{Quark Matter in Massive Compact Stars}, \emph{Astrophys. J.} {\bf 740} (2011) L14.
\bibitem{Bhattacharyya2016} S. Bhattacharyya, I. Bombaci, D. Logoteta and A.V. Thampan, \emph{Fast spinning strange stars: possible ways to constrain interacting quark matter parameters}, \emph{Mon. Not. R. Astron. Soc.} {\bf 457} (2016) 3101.
\bibitem{Zhou2018} E.-P. Zhou, X. Zhou and A. Li, \emph{Constraints on interquark interaction parameters with GW170817 in a binary strange star scenario}, \emph{Phys. Rev. D} {\bf 97} (2018) 083015.
\bibitem{Wang2019} Y.-B. Wang, X. Zhou, N. Wang and X.-W. Liu, \emph{The r-mode instability windows of strange stars}, \emph{Res. Astron. Astrophys.} {\bf 19} (2019) 30.
\bibitem{Workman2022} Particle Data Group, \emph{Review of Particle Physics}, \emph{Prog. Theor. Exp. Phys.} {\bf 2022} (2022) 083C01.
\bibitem{Fraga2001} E.S. Fraga, R.D. Pisarski and J. Schaffner-Bielich, \emph{Small, dense quark stars from perturbative QCD}, \emph{Phys. Rev. D} {\bf 63} (2001) 121702.
\bibitem{Alford2005} M. Alford, M. Braby, M. Paris and S. Reddy, \emph{Hybrid Stars that Masquerade as Neutron Stars}, \emph{Astrophys. J.} {\bf 629} (2005) 969.
\bibitem{Alford2001} M. Alford, K. Rajagopal, S. Reddy and F. Wilczek, \emph{Minimal color-flavor-locked-nuclear interface}, \emph{Phys. Rev. D} {\bf 64} (2001) 074017.
\bibitem{Rajagopal2001} K. Rajagopal and F. Wilczek, \emph{Enforced Electrical Neutrality of the Color-Flavor Locked Phase}, \emph{Phys. Rev. Lett.} {\bf 86} (2001) 3492.

\bibitem{Lee1956} T.D. Lee and C.N. Yang, \emph{Question of Parity Conservation in Weak Interactions}, \emph{Phys. Rev.} {\bf 104} (1956) 254.
\bibitem{Kobzarev1966} I.Y. Kobzarev, L.B. Okun and I.Y. Pomeranchuk, \emph{On the possibility of experimental observation of mirror particles}, \emph{Sov. J. Nucl. Phys.} {\bf 3} (1966) 837.

\bibitem{Blinnikov1982} S.I. Blinnikov and M.Y. Khlopov, \emph{On possible effects of 'mirror' particles}, \emph{Sov. J. Nucl. Phys.} {\bf 36} (1982) 472.
\bibitem{Blinnikov1983} S.I. Blinnikov and M.Y. Khlopov, \emph{Possible Astronomical Effects of Mirror Particles}, \emph{ Sov. Astron.} {\bf 27} (1983) 371.

\bibitem{Khlopov1991} M.Y. Khlopov, G.M. Beskin, N.G. Bochkarev, L.A. Pustilnik and S.A. Pustilnik, \emph{Observational Physics of the Mirror World}, \emph{Sov. Astron.} {\bf 35} (1991) 21.
\bibitem{Foot1991} R. Foot, H. Lew and R.R. Volkas, \emph{A model with fundamental improper spacetime symmetries}, \emph{Phys. Lett. B} {\bf 272} (1991) 67.

\bibitem{Foot2004} R. Foot, \emph{Mirror Matter-Type Dark Matter}, \emph{Int. J. Mod. Phys. D} {\bf 13} (2004) 2161.
\bibitem{Berezhiani2004} Z. Berezhiani, \emph{Mirror World and its Cosmological Consequences}, \emph{Int. J. Mod. Phys. A} {\bf 19} (2004) 3775.
\bibitem{BEREZHIANI2005} Z. Berezhiani, \emph{Through the Looking-Glass Alice's Adventures in Mirror World}, in \emph{From Fields to Strings: Circumnavigating Theoretical
Physics}, World Scientific, Singapore (2005) 2147.
\bibitem{Okun2007} L.B. Okun, \emph{Mirror particles and mirror matter: 50 years of speculation and searching}, \emph{Phys. Usp.} {\bf 50} (2007) 380.
\bibitem{Foot2014} R. Foot, \emph{Mirror dark matter: Cosmology, galaxy structure and direct detection}, \emph{Int. J. Mod. Phys. A} {\bf 29} (2014) 1430013.
\bibitem{Berezhiani2018} Z. Berezhiani, \emph{Matter, dark matter, and antimatter in our Universe}, \emph{Int. J. Mod. Phys. A} {\bf 33} (2018) 1844034.
\bibitem{Pavsic1974} M. Pavsic, \emph{External inversion, internal inversion, and reflection invariance}, \emph{Int. J. Theor. Phys.} {\bf 9} (1974) 229.

 \bibitem{Berezhiani2006} Z. Berezhiani and L. Bento, \emph{Neutron-Mirror-Neutron Oscillations: How Fast Might They Be?} \emph{Phys. Rev. Lett.} {\bf 96} (2006) 081801.
 
 \bibitem{Berezhiani2009} Z. Berezhiani, \emph{More about neutron-mirror neutron oscillation}, \emph{Eur. Phys. J. C} {\bf 64} (2009) 421.

 \bibitem{Goldman2019} I. Goldman, R.N. Mohapatra and S. Nussinov, \emph{Bounds on neutron-mirror neutron mixing from pulsar timing}, \emph{Phys. Rev. D} {\bf 100} (2019) 123021.
. 
 \bibitem{McKeen2021} D. McKeen, M. Pospelov and N. Raj, \emph{Neutron Star Internal Heating Constraints on Mirror Matter}, \emph{Phys. Rev. Lett.} {\bf 127} (2021) 061805.

 \bibitem{Goldman2022} I. Goldman, R.N. Mohapatra, S. Nussinov and  Y. Zhang, \emph{Neutron-Mirror-Neutron Oscillation and Neutron Star Cooling}, \emph{Phys. Rev. Lett.} {\bf 129} (2022) 061103. 

\bibitem{Capano2020} C.D. Capano et al., \emph{Stringent constraints on neutron-star radii from multimessenger observations and nuclear theory}, \emph{Nat. Astron.} {\bf 4} (2020) 625.
\bibitem{Riley2019} T.E. Riley et al., \emph{A NICER View of PSR J0030+0451: Millisecond Pulsar Parameter Estimation}, \emph{Astrophys. J. Lett.} {\bf 887} (2019) L21.
\bibitem{Miller2019} M.C. Miller et al., \emph{PSR J0030+0451 Mass and Radius from NICER Data and Implications for the Properties of Neutron Star Matter}, \emph{Astrophys. J. Lett.} {\bf 887} (2019) L24.

\bibitem{Riley2021} T.E. Riley et al., \emph{A NICER View of the Massive Pulsar PSR J0740+6620 Informed by Radio Timing and XMM-Newton Spectroscopy}, \emph{Astrophys. J. Lett.} {\bf 918} (2021) L27.
\bibitem{Miller2021} M.C. Miller et al., \emph{The Radius of PSR J0740+6620 from NICER and XMM-Newton Data}, \emph{Astrophys. J. Lett.} {\bf 918} (2021) L28.

\bibitem{Schaab1997} C. Schaab, B. Hermann, F. Weber and M.K. Weigel, \emph{Are strange stars distinguishable from neutron stars by their cooling behaviour?}, \emph{J. Phys. G: Nucl. Part. Phys.} {\bf 23} (1997) 2029.
\bibitem{Pi2015} C.-M. Pi, S.-H. Yang and X.-P. Zheng, \emph{R-mode instability of strange stars and observations of neutron stars in LMXBs}, \emph{Res. Astron. Astrophys.} {\bf 15} (2015) 871.
\bibitem{Yang2020} S.-H. Yang, C.-M. Pi, X.-P. Zheng and F. Weber, \emph{Non-Newtonian Gravity in Strange Quark Stars and Constraints from the Observations of PSR J0740+6620 and GW170817}, \emph{Astrophys. J.} {\bf 902} (2020) 32.
\bibitem{Yang2021a} S.-H. Yang, C.-M. Pi, X.-P. Zheng and F. Weber, \emph{Constraints from compact star observations on non-Newtonian gravity in strange stars based on a density dependent quark mass model}, \emph{Phys. Rev. D} {\bf 103} (2021) 043012.
\bibitem{Pi2022} C.-M. Pi and S.-H. Yang, \emph{Non-Newtonian gravity in strange stars and constraints from the observations of compact stars}, \emph{New Astron.} {\bf 90} (2022) 101670.
\bibitem{Ma2023} Z.-J. Ma, Z.-Y. Lu, J.-F. Xu, G.-X. Peng, X. Fu and J. Wang, \emph{Cold quark matter in a quasiparticle model: Thermodynamic consistency and stellar properties}, \emph{Phys. Rev. D} {\bf 108} (2023) 054017.

\bibitem{Baym2018} G. Baym, T. Hatsuda, T. Kojo, P.D. Powell, Y. Song and T. Takatsuka, \emph{From hadrons to quarks in neutron stars: a review}, \emph{ Rept. Prog. Phys.} {\bf 81} (2018) 056902. 

\bibitem{Leonhardt2020} M. Leonhardt, M. Pospiech, B. Schallmo, J. Braun, C. Drischler, K. Hebeler and A. Schwenk, \emph{Symmetric Nuclear Matter from the Strong Interaction}, \emph{Phys. Rev. Lett.} {\bf 125} (2020) 142502. 
\bibitem{Alford2003} M. Alford and S. Reddy, \emph{Compact stars with color superconducting quark matter}, \emph{Phys. Rev. D} {\bf 67} (2003) 074024. 

\bibitem{Hippert2022} M. Hippert, J. Setford, H. Tan, D. Curtin, J. Noronha-Hostler and N. Yunes, \emph{Mirror neutron stars}, \emph{Phys. Rev. D} {\bf 106} (2022) 035025. 






\end{thebibliography}
\end{document}